\documentclass[paper, nofootinbib]{revtex4}
\usepackage{amsfonts}
\usepackage{graphicx}
\usepackage[latin1]{inputenc}
\usepackage{amsmath}
\usepackage{amssymb}

\begin{document}


\title{Frenkel electron on an arbitrary electromagnetic background and magnetic {\it \bf Zitterbewegung}
}
\author{Alexei A. Deriglazov}
\email{alexei.deriglazov@ufjf.edu.br}
\affiliation{%
Depto. de Matem\'atica, ICE, Universidade Federal de Juiz de Fora, MG, Brasil \\ and \\ Laboratory of Mathematical
Physics, Tomsk Polytechnic University, 634050 Tomsk, Lenin Ave. 30, Russian Federation}

\author{Andrey M. Pupasov-Maksimov}
 \email{pupasov@phys.tsu.ru}
\affiliation{%
Depto. de Matem\'atica, ICE,\\
Universidade Federal de Juiz de Fora, MG, Brasil
}%
\date{\today}

\begin{abstract}
We present Lagrangian which implies both necessary constraints and dynamical equations for position and spin of
relativistic spin one-half particle. The model is consistent for any value of magnetic moment $\mu$ and for arbitrary
electromagnetic background. Our equations coincide with those of Frenkel in the approximation in which the latter have
been obtained by Frenkel. Transition from approximate to exact equations yields two structural modifications of the
theory. First, Frenkel condition on spin-tensor turns into the Pirani condition. Second, canonical momentum is no more
proportional to velocity. Due to this, even when $\mu=1$ (Frenkel case), the complete and approximate equations predict
different behavior of particle. The difference between momentum and velocity means extra contribution into
spin-orbit interaction. To estimate the contribution, we found exact solution to complete equations for the case of
uniform magnetic field. While Frenkel electron moves around the circle, our particle experiences magnetic {\it
Zitterbewegung}, that is oscillates in the direction of magnetic field with amplitude of order of Compton wavelength
for the fast particle. Besides, the particle has dipole electric moment.
\end{abstract}

\keywords{Semiclassical Description of Relativistic Spin, Dirac Equation, Theories with Constraints}

\maketitle

\section{Introduction \label{intro}}

Consistent and complete description of spin effects of the relativistic electron is achieved in QED on the base of
Dirac equation. However, starting from the pioneer works \cite{pryce1948mass, newton1949localized, foldy:1978} and up
to present date, interpretation of final results in some cases is under permanent and controversial debates in various
theoretical and experimental set-ups \cite{vanHolten(1991), kidder:1993, khriplovich:1996, khriplovich:1998,
pomeransky:1999, Chakrabarti(2000), obukhov:2011, Jaffe:1990, Wakamatsu:2010, selva:96, rafanelli-collinearity,
Zahringer:2010, AAD6}. Understanding of spin precession in the case of arbitrary magnetic moment in an external
electromagnetic field is important in the development of experimental technics for measurements of anomalous magnetic
moment \cite{Field:1979, miller2007muon}. In accelerator physics \cite{hoffstaetter2006adiabatic} it is important to
control resonances leading to depolarization of a beam. In the case of vertex electrons carrying arbitrary angular
momentum, semiclassical description can also be useful \cite{karlovets2012}. So the relationship among classical and
quantum descriptions remains an important step of analysis, providing the interpretation of results of QFT computations
in usual terms: particles and their interactions. Hence an actual task is to develop, in a systematic form, the
classical model of an electron \cite{Frenkel, Frenkel2, BMT, corben:1968, hanson1974relativistic, berezin:1977,
Marnelius(1990), Gitman(1997), balachandran:1980, bt, bbra, grassberger1978, cognola1981lagrangian, Marcus(1995),
Stern(1987), Vacaru(2002), Alexei} which would be as close as possible to the Dirac equation.

Maybe the best candidates for classical equations of relativistic electron are those of Frenkel \cite{Frenkel,
Frenkel2} and Bargmann, Michel and Telegdi (BMT) \cite{BMT}. They almost exactly reproduce spin dynamics of polarized
beams in uniform fields, and thus might be proper classical analog for the Dirac theory. However, to be able to
describe other spin effects, it is desirable to have systematically constructed Lagrangian and Hamiltonian formulations
as well as proper quantization scheme for these equations (note that one needs a Hamiltonian to describe, for instance,
Stark and Zeeman effects). Then it would be possible to use them as a semiclassical approximation of the QFT
computations based on Dirac equation.

Non relativistic spin operators are proportional to the Pauli matrices, so they form a simple algebra with respect to
commutator
\begin{eqnarray}\label{intr.05}
[\hat S_i, \hat S_j]_{-}=i\hbar\epsilon_{ijk}\hat S_k,
\end{eqnarray}
as well as with respect to anticommutator
\begin{eqnarray}\label{intr.06}
[\hat S_i, \hat S_j]_{+}=\frac{\hbar^2}{2}\delta_{ij}.
\end{eqnarray}
These equations prompt that spin-space in classical model can be described by either even or odd (Grassmann) variables.
The pioneer model based on odd variables have been constructed by Berezin and Marinov \cite{berezin:1977}. This gives
very economic and elegant scheme for semiclassical description of spin. For non relativistic spin, the Lagrangian reads
$\frac{m}{2}(\dot x_i)^2+i\xi_i\dot\xi_i$, where the inner space of spin is constructed from vector-like Grassmann
variables $\xi_i$, $\xi_i\xi_j=-\xi_j\xi_i$. Since the Lagrangian is linear on $\dot\xi_i$, their conjugate momenta
coincide with $\xi$, $\pi_i=\frac{\partial L}{\partial\dot\xi_i}=i\xi_i$. The relations represent the Dirac
second-class constraints and are taken into account by transition from the grassmannian Poisson bracket to the Dirac
one. After that, the constraints can be used to exclude $\pi_i$. Dirac bracket of the remaining variables reads
$\{\xi_i, \xi_j\}_{DB}=i\delta_{ij}$.
Comparing this with Eq. (\ref{intr.06}), we quantize the model replacing $\xi_i\rightarrow\frac{\hbar}{2}\sigma_i$.
Relativistic spin is described in a similar way \cite{berezin:1977, Galvao80, Marnelius(1990), Gitman(1997)}. The
problem here is that Grassmann classical mechanics represents a rather formal mathematical construction. It leads to
certain difficulties \cite{berezin:1977, Gitman(1997)} in attempts to use it for description the spin effects on the
semiclassical level, before the quantization. Besides, generalization of Grassmann mechanics to higher spins is not
known \cite{Gitman:1999}. Hence it would be interesting to describe spin on a base of usual variables, that is we
intend to arrive at the commutator algebra (\ref{intr.05}) instead of (\ref{intr.06}).

Contrary to the models based on commuting spinors \cite{bt, bbra}, in the Berezin-Marinov approach the $\sigma_i$ (or
$\gamma^\mu$) matrices do not appear in classical theory but produced through the quantization process. The same turns
out to be true in our model based on non-Grassmann vector for description of spin.

Very general approach to description of rotational degrees of freedom in relativistic theory has been developed by
Hanson and Regge \cite{hanson1974relativistic}. They suggested to represent a relativistic spherical top as a point on
a world-line to which a body-fixed frame is attached. The frame is identified with the Lorentz-group element, so the
trajectory $(x^\mu(\tau), \Lambda^{\mu\nu}(\tau))$ of the top represents a line of the Poincare-group manifold. The
antisymmetric tensor $\Lambda^{-1}\dot\Lambda$ has been chosen as the basic quantity to describe the rotational degrees
of freedom. They asked on the most general form of Lagrangian which yields only three physical rotational degrees of
freedom. The Lagrangian gives generalized mass-shell constraint which relates mass with spin, so in quantum mechanics
they obtained a string-like spectrum composed by a family of particles with varying mass and spin. They also analyzed
whether their spin-tensor couples directly with electromagnetic fields, and concluded on impossibility to construct the
interaction in a closed form. As we show below, this can be achieved in closed form for the vector that constitutes our
spin-tensor.

Since the commutator (\ref{intr.05}) represents the angular-momentum algebra, it is natural to represent the spin in
classical theory as the composite quantity, $\vec S=\vec\omega\times\vec\pi$, constructed from spacial components of
some inner-space coordinate $\omega^\mu$ and its conjugated momentum $\pi^\mu$. As in the Hanson-Regge approach, the
main problem here is to construct the Poincare-invariant Lagrangian which has the right number of degrees of freedom
and admits an interaction in closed and relatively simple form. We need a variational problem which yields the
appropriate constraints. In turn, this implies the use of Dirac's machinery for analysis of constrained systems. Though
a number of models \cite{balachandran:1980, bt, bbra, grassberger1978, cognola1981lagrangian, Kosyakov:2003} with
vector variables yield Frenkel and BMT equations, they also contain extra degrees of freedom. At the classical level
one can simply ignore them. However, they should be taken into account during quantization procedure, this leads to
quantum models essentially different from the Dirac electron.

In this work we continue detailed analysis of the Frenkel and BMT equations started in \cite{DPM1, DPM2}, and construct
the Lagrangian which yields generalization of these equations to the case of arbitrary electromagnetic background. Even
for non interacting theory search for Lagrangian represents rather non trivial problem \cite{Frenkel2,
hanson1974relativistic}. In \cite{DPM2} we have solved this problem, considering spin as angular momentum of inner
four-dimensional vector space attached to the point of a world-line.

In Hamiltonian formulation the model represents a nontrivial example of a constrained system. Phase space of the model
turns out to be curved manifold equipped, in a natural way, with the structure of fiber bundle. Detailed analysis of
the underlying geometry has been presented in \cite{DPM1}. This allowed us to develop the proper quantization scheme.
In \cite{DPM2} we have performed both canonical (in physical-time parametrization) and manifestly covariant (in
arbitrary parametrization) quantization of the free model, and established the relation with one-particle sector of
Dirac equation as well as with quantum theory of two-component Klein-Gordon equation developed by Feynmann and
Gell-Mann \cite{feynman1958fermi-interaction}. It has been demonstrated that various known in the literature
non-covariant, covariant and manifestly-covariant operators of position and spin acquire clear meaning and
interpretation in the Lagrangian model of Frenkel electron. In particular, we have found the manifestly covariant form
of position and spin operators in the space of positive-energy Dirac spinors.

In the Hamiltonian formulation two second-class constraints appeared which, at the end,  supply the Frenkel condition
on spin-tensor. They depend on both position and spin-sector variables. This leads to new properties as compared with
non relativistic spin \cite{DPM1}. The constraints must be taken into account by transition from Poisson to Dirac
bracket, this leads to nonvanishing classical brackets for the position variables. In the result, the position space is
endowed, in a natural way, with noncommutative structure which originates from accounting of spin degrees of freedom.
Our model represents an example of a situation, when physically interesting noncommutative relativistic particle
emerges in a natural way. For the case, the "parameter of non-commutativity" is proportional to spin-tensor. As a
consequence, operators corresponding to position of the electron are non-commutative (they can be identified
\cite{DPM2} with Pryce (d) operators). This implies that effects of non-commutativity could be presented at the Compton
wave length, in contrast to conventional expectations of non-commutativity at Planck length.

There are a lot of candidates for spin and position operators of the relativistic electron \cite{pryce1948mass,
newton1949localized, foldy:1978, feynman1961, corben:1968}. Different position observables coincide when we consider
standard quasi-classical limit. So, in absence of a systematically constructed classical model of an electron it is
difficult to understand the difference between these operators. Our approach allows us to do this, after realizing all
them at the classical level. Besides, all the candidates obey the same equations in free theory, so the question of
which of them are the true position and spin is a matter of convention. The situation changes in the interacting theory
considered below, where we can distinguish the variables according their classical dynamics in an external field.

In the present work we construct and study an interacting theory. In section \ref{subsec:interactiion-L-H}  we show
that our Lagrangian admits interaction with an arbitrary electromagnetic background\footnote{Interaction with an
arbitrary curved background is presented in \cite{DPW2}.}. The model contains two coupling constants - charge $e$ and
the interaction constant $\mu$ of basic spin variables with $F_{\mu\nu}$. Provisionally, we call this magnetic moment.
The theory is consistent for arbitrary values of $\mu$. For the position variable we have the minimal interaction term
$\frac{e}{c}A_\mu\dot{x}^\mu$. As for spin, when the particle has non-vanishing magnetic moment, this interacts with
electromagnetic field in a highly nonlinear way. This turns out to be necessary for preservation of the number and
algebraic structure of constraints in the passage from free to interacting theory. In section
\ref{subsec:interactiion-eqs} we present and analyze equations of motion in Hamiltonian and Lagrangian formulations. We
show that they follow from simple and expected Hamiltonian (\ref{phys.ham}), when we deal with the Dirac bracket.  We
compare our equations with those of Frenkel \cite{Frenkel, Frenkel2}. Frekel considered the case $\mu=1$, and found his
equations in the quadratic approximation on spin-tensor. We show that our exact equations coincide with those of
Frenkel in these limits. Hence our Lagrangian gives complete Frenkel equations for arbitrary field and magnetic moment.

Frenkel tensor can be used to construct BMT-type four-vector. We write the corresponding equations of motion. While
Hamiltonian equations can be rewritten in closed form in terms of BMT vector, see Eq. (\ref{uf4.14})-(\ref{uf4.15}), we
do not achieved this for Lagrangian equations in our theory. In the Lagrangian form, the equation for BMT vector
contains Frenkel tensor, see Eq. (\ref{pp16}). It seems that Frenkel spin in our theory represents more fundamental
object as compared with BMT spin.

While equations of motion have a rather complicated structure, in the case of uniform magnetic field there are a lot of
symmetries and hence integrals of motions providing complete analytical solution. In section
\ref{sec:interaction-example} we find exact solution to our equations for this case. As compared with Frenkel and BMT
equations, our model takes into account two effects. First, magnetic moment interacted with a magnetic field results in
additional mass of electron. Second, in the case of anomalous magnetic moment the velocity and the momentum are not
collinear, this modifies the Lorentz force. Our model naturally incorporates both these effects and leads to small
corrections of the trajectory and spin precession.

\section{Lagrangian and Hamiltonian of interacting theory \label{subsec:interactiion-L-H}}
To start with, we shortly describe the structure of free theory \cite{DPM1, DPM2}. Configuration space of the model
consist of the position $x^\mu(\tau)=(ct, {\boldsymbol x})$ and the vector-like variable of spin
$\omega^\mu(\tau)=(\omega^0, {\boldsymbol\omega})$, ${\boldsymbol{\omega}}=(\omega_1, \omega_2, \omega_3)$ in an
arbitrary parametrization $\tau$. $p^\mu$  and $\pi^\mu$ are conjugate momenta for $x^\mu$ and $\omega^\mu$. The
variables in the free theory are subject to the constraints (we set $a_3=\frac{3\hbar^2}{4a_4}$)
\begin{eqnarray}\label{1.1}
T_1=p^2+(mc)^2=0\,, \qquad \qquad \qquad \qquad \qquad \qquad \\
T_3=\pi^2-a_3=0\,, \quad T_4=\omega^2-a_4=0\,, \quad
T_5=\omega\pi=0\,, \label{1.2}\\
T_6=   p\omega=0\,, \qquad T_7=p\pi=0\,, \qquad \qquad \qquad \qquad \label{1.3}
\end{eqnarray}
where $\omega\pi=-\omega^0\pi^0+{\boldsymbol\omega\boldsymbol\pi}$ and so on. As the Hamiltonian action functional, we
simply take $L_H=p\dot x+\pi\dot\omega-H$, with the Hamiltonian $H=g_iT_i$ in the form of linear combination the
constraints $T_i$ multiplied by auxiliary variables $g_i$, $i=1, 3, 4, 5, 6, 7$. The constraint $T_3$ belongs to
first-class and is related with local spin-plane symmetry presented in the theory \cite{DPM1}. The basic spin-sector
variables change under the symmetry, so they do not represent an observables quantities. As the observable quantity we
take the Frenkel spin-tensor $J^{\mu\nu}$
\begin{eqnarray}\label{1.4}
J^{\mu\nu}=2(\omega^\mu\pi^\nu-\omega^\nu\pi^\mu).
\end{eqnarray}
The constraints (\ref{1.2}) and (\ref{1.3}) imply the following restrictions (in the free theory the conjugated
momentum is proportional to four-velocity, $p^\mu\sim u^\mu$)
\begin{eqnarray}\label{1.5}
J^{\mu\nu}p_\nu=0, \quad J^{\mu\nu}J_{\mu\nu}=6\hbar^2.
\end{eqnarray}
Spacial components of the Frenkel tensor can be used to construct the quantity
\begin{eqnarray}\label{spin}
S^i=\frac{1}{4}\epsilon^{ijk}J_{jk}\,,
\end{eqnarray}
which we identify with non relativistic spin of Pauli theory. In the interacting theory $p^\mu$ turns into canonical
momentum ${\cal P}^\mu$, which for non uniform fields or/and $\mu\ne 1$ does not proportional to four-velocity. Hence
in this case the Frenkel condition $J^{\mu\nu}u_\nu=0$ turns into the Pirani condition \cite{pirani:1956,
tulczyjew:1959, dixon:1964} $J^{\mu\nu}{\cal P}_\nu=0$.

Frenkel tensor can be used to construct four-vector
\begin{eqnarray}\label{1.6}
s^\mu(\tau)\equiv\frac{1}{4\sqrt{-p^2}}\epsilon^{\mu\nu\alpha\beta}p_\nu J_{\alpha\beta}, \quad \mbox{then} \quad s^\mu
p_\mu=0, \quad s^2=\frac{3\hbar^2}{4}.
\end{eqnarray}
In our theory, even in the case of interaction, the condition $s^\mu{\cal P}_\mu=0$ implies $s^\mu u_\mu=0$, see Eq.
(\ref{uf4.121}) below. So we can identify $s^\mu$ with BMT vector \cite{BMT}. In the rest frame, spacial components
$s^i$ of BMT vector coincide with $S^i$. In an arbitrary frame, they are related as follows:
\begin{eqnarray}\label{spin.1}
S^i=\frac{p^0}{\sqrt{-p^2}}\left(\delta_{ij}-\frac{p_ip_j}{(p^0)^2}\right)s^j\,.
\end{eqnarray}
In the free theory the equation (\ref{1.6}) can be inverted,
$J^{\mu\nu}=-\frac{2}{\sqrt{-p^2}}\epsilon^{\mu\nu\alpha\beta}p_\alpha s_\beta$, so the two quantities are
mathematically equivalent. In the interacting theory, we have $p\rightarrow{\cal P}={\cal P}(u, J, F)$, see Eqs.
(\ref{uf4.7}) and (\ref{uf4.71}), so (\ref{1.6}) becomes non linear equation.

In \cite{DPM2} we developed Lagrangian formulation of the theory. Excluding conjugate momenta from $L_H$, we obtained
the Lagrangian action. Further, excluding the auxiliary variables, one after another, we obtained various equivalent
formulations of the model. In the end, we got the "minimal" formulation without auxiliary variables. This
reads\footnote{The last term in (\ref{intr.5}) represents kinematic (velocity-independent) constraint which is well
known from classical mechanics. So, we might follow the classical-mechanics prescription to exclude $g_4$ as well. But
this would lead to lose of manifest covariance of the formalism.}
\begin{eqnarray}\label{intr.5}
S=\int d\tau -mc\sqrt{-\dot xN\dot x}+\sqrt{a_3}\sqrt{\dot\omega N\dot\omega}-\frac{1}{2}g_4(\omega^2-a_4),
\end{eqnarray}
where $N^{\mu\nu}\equiv\eta^{\mu\nu}-\frac{\omega^\mu\omega^\nu}{\omega^2}$ is projector on the plane transverse to the
direction of $\omega^\mu$. The action is written in a parametrization $\tau$ which obeys $\frac{dt}{d\tau}>0$.

Interaction with an external background should not spoil the number and algebraic properties of constraints
(\ref{1.1})-(\ref{1.3}). We do not know how to achieve this for the minimal action\footnote{Hanson and Regge
\cite{hanson1974relativistic} have found highly non linear interaction for the case of their relativistic top. It would
be interesting to apply their formalism to our minimal action.}. For instance, the natural reparametrization-invariant
interaction $\frac{e}{c}A_\mu\dot x^\mu+\mu F_{\mu\nu}\omega^\mu\dot\omega^\nu$, even for vanishing magnetic moment,
leads to the theory with the number and algebraic structure of constraints different from those of free theory. So we
start with the equivalent Lagrangian with four auxiliary variables, $g_1, g_3, g_4$ and $g_7$, this turns out to be
appropriate to our aims. Of course, the auxiliary variables will be excluded from final equations of motion, see Eqs.
(\ref{pp11})-(\ref{pp11-J-F0}).

To introduce coupling of the position variable with an electro-magnetic field, we add the minimal interaction term
$A_\mu\dot x^\mu$. As for spin, we propose to modify derivative of $\omega$ as follows
\begin{eqnarray}\label{il1}
\dot\omega^\mu ~ \rightarrow ~ D\omega^\mu=\dot\omega^\mu-g_1\frac{e\mu}{c}(F \omega)^\mu\,.
\end{eqnarray}
This is the only term which we have found to be consistent with the constraints $T_i$. Lagrangian reads
\begin{eqnarray}\label{lagrangian-bmt-em-4aux-vars}
L=\frac{1}{2\det\tilde G}\left[ g_3\left(\dot xN\dot x\right)-2g_7\left(\dot xND\omega\right) +g_1\left(D\omega
ND\omega\right) \right]+\frac{e}{c}A_\mu\dot x^\mu- \cr
\frac{g_4}{2}(\omega^2-a_4)-\frac{g_1}{2}m^2c^2+\frac{g_3}{2}a_3 \,,
\end{eqnarray}
where $\mbox{det}\tilde G=g_1g_3-g_7^2$.  Since $L$ contains auxiliary variables, even for $\mu=0$ we have highly
nonlinear interaction. As a consequence, motion of spin influences motion of the particle and {\it vice versa}.

We first establish whether our Lagrangian gives the desired constraints. The momenta read
\begin{eqnarray}\label{def-lagr-momentumP}
p^\mu &=&\frac{\partial L}{\partial \dot{x}^\mu}=\frac{1}{\det\tilde
G}\left(g_3N\dot{x}^\mu-g_7ND\omega^\mu\right)+\frac{e}{c}A^\mu\,, \cr \pi^\mu &=& \frac{\partial L}{\partial
\dot{\omega}^\mu}=\frac{1}{\det\tilde G}\left(-g_7N\dot{x}^\mu+g_1ND\omega^\mu\right)\,,
\end{eqnarray}
\begin{equation}\label{il2}
\pi_{g_i}=\frac{\partial L}{\partial \dot{g}_i}=0\,.
\end{equation}
According to (\ref{il2}), the momenta $\pi_{gi}$ represent primary constraints, $\pi_{gi}=0$. Using the property
$N\omega=0$ of the projector $N$, from Eqs. (\ref{def-lagr-momentumP}) more primary constraints follow,
$T_5=\pi\omega=0$ and $T_6={\cal P}\omega=0$. It has been denoted
\begin{equation}\label{il3}
{\cal P}^\mu=p^\mu-\frac{e}{c}A^\mu\,.
\end{equation}
To write Hamiltonian, we solve the system (\ref{def-lagr-momentumP}) with respect to projected velocities
\begin{equation}
N\dot{x}^\mu=g_1{\cal P}^\mu+g_7\pi^\mu\,, \qquad ND\omega^\mu=g_3\pi^\mu+g_7{\cal P}^\mu\,.
\end{equation}
Using these expressions as well as the identities ${\cal P}\dot{x}={\cal P}N\dot{x}$, $\pi\dot{\omega}=\pi N
\dot{\omega}$ we obtain Hamiltonian $H=p\dot x+\pi\dot\omega-L+\lambda_a\Phi_a$ in the form
\begin{eqnarray}\label{bmt-em-hamiltonian-expl}\label{il3.1}
H=\frac{g_1}{2}\left({\cal{P}}^2-\frac{\mu e}{2c}(JF)+m^2c^2\right) +\frac{g_3}{2}\left(\pi^2-a_3\right)+
\frac{g_4}{2}\left(\omega^2-a_4\right) + \cr \lambda_5\left(\omega\pi\right)+ \lambda_6\left(\cal{P}\omega\right) +g_7
\left(\cal{P}\pi\right) + \lambda_{g_i}\pi_{g_i}\,,
\end{eqnarray}
where $\lambda_5$ and $\lambda_6$ appear as Lagrangian multipliers for primary constraints $T_5$ and $T_6$. We denote
$(JF)=J^{\mu\nu}F_{\mu\nu}$ and so on. From (\ref{bmt-em-hamiltonian-expl}) we conclude that $T_1$, $T_3$, $T_4$ and
$T_7$ appear as secondary constraints when we impose the compatibility conditions $\dot{\pi}_{g_i}=\{\pi_{gi}, H\}=0$.
The second, third and fourth stages of the Dirac-Bergmann algorithm can be resumed as follows
\begin{eqnarray}\label{third-stage-L4toHam}
T_1=0 ~ \quad &\Rightarrow& \quad  \lambda_6C+ g_7D=0\,,\label{il2.1} \\
T_3=0 ~ \quad &\Rightarrow& \quad ~  \lambda_5=0\,,\label{il2.2} \\
T_4=0 ~ \quad &\Rightarrow&\quad ~  \lambda_5=0\,,\label{il2.3} \\
T_5=0 ~ \quad &\Rightarrow&\quad ~ g_4=\frac{a_3}{a_4}g_3\,, ~ \qquad \quad \Rightarrow\quad ~
\lambda_{g_4}=\frac{a_3}{a_4}\lambda_{g3}\,,\label{il2.4} \\
T_6=0 ~ \quad &\Rightarrow&\quad ~ g_1C-g_7 M^2c^2=0\,, \quad\Rightarrow\quad ~ \lambda_{g7}=f(\lambda_{g1})\,,\label{il2.5} \\
T_7=0 ~ \quad &\Rightarrow&\quad ~ g_1D+\lambda_6 M^2c^2=0\,. ~ \label{il2.6}
\end{eqnarray}
We have denoted
\begin{eqnarray}\label{il0}
M^2=m^2-\frac{e(2\mu+1)}{4c^3}F_{\mu\nu}J^{\mu\nu},
\end{eqnarray}
\begin{eqnarray}\label{il4}
C=-\frac{e}{c}(\mu-1)(\omega F{\cal P})+\frac{e\mu}{4c}(\omega\partial)(JF), \cr D=-\frac{e}{c}(\mu-1)(\pi F{\cal
P})+\frac{e\mu}{4c}(\pi\partial)(JF).
\end{eqnarray}
Eq. (\ref{il2.1}) turns out to be a consequence of (\ref{il2.5}) and (\ref{il2.6}),
$\lambda_6(\ref{il2.5})+g_7(\ref{il2.6})=g_1(\ref{il2.1})$, and can be omitted. Eq. (\ref{il2.5}) determines
$g_7=\frac{C}{M^2c^2}g_1$ while (\ref{il2.6}) gives the lagrangian multiplier $\lambda_6=-\frac{D}{M^2c^2}g_1$. The
Dirac-Bergmann algorithm stops at the fourth stage. This yields all the desired constraints $T_a$, $a=1, 3, 4, 5, 6,
7$. Two auxiliary variables, $g_1$ and $g_3$, and the corresponding Lagrange multipliers $\lambda_{g_1}$,
$\lambda_{g_3}$ have not been determined.

It is useful to summarize the algebra of Poisson brackets between constraints in a compact form, see Table
\ref{tabular:algebra-constraints-BMT-em}.
\begin{table}
\caption{Algebra of constraints} \label{tabular:algebra-constraints-BMT-em}
\begin{center}
\begin{tabular}{c|c|c|c|c|c|c}
                              & $\qquad T_1 \qquad$  & $T_3$         & $T_4$          & $T_5$                 & $T_6$    & $T_7$     \\  \hline \hline
$T_1=\mathcal{P}^2- $         & 0             & 0             & 0              & 0                     & -2C   & -2D     \\
$\frac{\mu e}{2c}F^{\mu\nu}J_{\mu\nu}+m^2c^2$
                              &               &               &           &            &               &             \\
\hline
$T_3=\pi^2-a_3$               & 0             & 0             & $-4T_5$   & $-2(a_3+T_3)$          & $-2T_7$ & 0\\
& ${}$ &      &           &                &
&        \\
\hline
$T_4=\omega^2-a_4$            & 0             & $4T_5$  & $0$            & $2(T_4+a_4)$          & 0       & $2T_6$\\
& ${}$ &      &           &                &
&        \\
\hline
$T_5=\omega\pi$               & 0     & $2(T_3+a_3)$      & $-2(a_4+T_4)$   &     0                 & $-T_6$  & $T_7$\\
& ${}$ &      &           &                &
&        \\
\hline
$T_6=\mathcal{P}\omega$       & $2C$  &$2T_7$  & $0$&   $T_6$               & 0       & $T_1-M^2c^2$  \\
                              & ${}$ &      &           &                &
&        \\
\hline
$T_7=\mathcal{P}\pi$          & $2D$  & 0         & $-2T_6$  & $-T_7$          &$-T_1+M^2c^2$& 0 \\
                              & ${}$ &       &           &                &
&        \\
\hline
\end{tabular}
\end{center}
\end{table}
We note that Poisson brackets of $T_1$ and $\tilde T_3=T_3+\frac{a_3}{a_4}T_4$ vanish on the constraint surface, so
they form the first-class subset. The presence of two first-class constraints is in a correspondence with the fact that
two lagrangian multipliers remain undetermined within the Dirac procedure. Matrix of Poisson brackets of the remaining
constraints, $T_4, T_5, T_6$ and $T_7$, is nondegenerate, so this is a set of second-class constraints. All this is in
correspondence with free theory \cite{DPM2}.

In resume, the interaction does not spoil the structure and algebraic properties of Hamiltonian constraints of the free
theory.

\section{Exact Frenkel equations on arbitrary background  \label{subsec:interactiion-eqs}}

\subsection{Hamiltonian equations of motion \label{subsec:interactiion-H-eqs}}

The Hamiltonian (\ref{il3.1}) determines evolution of the basic variables through the Poisson bracket $\dot q=\{q,
H\}$. Equivalently, we can pass from Poisson to Dirac bracket constructed on the base of the second-class constraints
$T_4, T_5, T_6, T_7$. The list of Dirac brackets is presented in the Appendix. After that, our highly nonlinear
interaction turns out to be hidden in the Dirac bracket: the constraints can be used in the Hamiltonian (\ref{il3.1}), this gives
the expression
\begin{eqnarray}\label{il3.11}
H_1=\frac{g_1}{2}\left({\cal{P}}^2-\frac{\mu e}{2c}(JF)+m^2c^2\right) +\frac{g_3}{2}\left(J^2-8a_3a_4\right)\,.
\end{eqnarray}
Equations of motion now can be obtained with help of $H_1$ and the Dirac bracket, $\dot q=\{q, H_1\}_{DB}$. They read
\begin{eqnarray}\label{uf4.5}
\dot x^\mu=g_1u^\mu, \qquad \dot{\cal P}^\mu=g_1\frac{e}{c}(Fu)^\mu+g_1\frac{\mu e}{4c}\partial^\mu(JF),
\end{eqnarray}
\begin{eqnarray}\label{uf4.6}
\dot\omega^\mu=g_1\frac{e\mu}{c}(F\omega)^\mu+g_3\pi^\mu+g_7{\cal P}^\mu, \quad \cr
\dot\pi^\mu=g_1\frac{e\mu}{c}(F\pi)^\mu-\frac{a_3}{a_4}g_3\omega^\mu-\lambda_6{\cal P}^\mu,
\end{eqnarray}
where $\partial^\mu(JF)=J^{\alpha\beta}\partial^\mu F_{\alpha\beta}$.  According to (\ref{il2.5}) and (\ref{il2.6}),
the four-velocity $u^\mu$ is not proportional to canonical momentum ${\cal P}^\mu$
\begin{eqnarray}\label{uf4.7}
u^\mu={\cal P}^\mu+\frac{g_7}{g_1}\pi^\mu+\frac{\lambda_6}{g_1}\omega^\mu=T^{\mu}{}_\nu{\cal P}^\nu+Y^\mu.
\end{eqnarray}
We have denoted
\begin{eqnarray}\label{uf4.71}
T^{\mu\nu}=\eta^{\mu\nu}-(\mu-1)a(JF)^{\mu\nu}, \qquad Y^\mu=\frac{\mu a}{4}J^{\mu\alpha}\partial_\alpha(JF)\,, \cr
a=-\frac{e}{2M^2c^3}\equiv\frac{-2e}{4m^2c^3-e(2\mu+1)(JF)}. \qquad \qquad
\end{eqnarray}
Matrix $T$ is invertible, the inverse matrix $\tilde T$ has the same structure (we used the identity
$(JFJ)^{\mu\nu}=-\frac12(JF)J^{\mu\nu}$ which implied by (\ref{1.4}))
\begin{eqnarray}\label{pp7}
\tilde T^{\mu\nu}=\eta^{\mu\nu}+(\mu-1)b(JF)^{\mu\nu}, \qquad \qquad \cr
b=\frac{2a}{2+(\mu-1)a(JF)}\equiv\frac{-2e}{4m^2c^3-3e\mu(JF)}.
\end{eqnarray}

All the basic variables have ambiguous evolution. $x^\mu$ and ${\cal P}^\mu$ have one-parametric ambiguity due to $g_1$
(they change under reparametrizations) while $\omega$ and $\pi$ have two-parametric ambiguity due to $g_1$ and $g_3$
(they change under reparametrizations and spin-plane symmetry). The quantities $x^\mu$, ${\cal P}^\mu$ and the
spin-tensor $J^{\mu\nu}$ are spin-plane invariants. Their equations of motion form a closed system
\begin{eqnarray}\label{uf4.8}
\dot x^\mu &=& g_1\left[{\cal P}^\mu-aJ^{\mu\alpha}\left((\mu-1)(F{\cal
P})_\alpha-\frac{\mu}{4}\partial_\alpha(JF)\right)\right],\\
\label{uf4.8-1}
\dot{\cal P}^\mu &=& \frac{e}{c}(F\dot x)^\mu+g_1\frac{\mu e}{4c}\partial^\mu(JF)\,,\\
\label{uf4.9} \dot J^{\mu\nu}&=&g_1\left[\frac{e\mu}{c}F^{[\mu}{}_\alpha J^{\alpha\nu]}-2a{\cal
P}^{[\mu}J^{\nu]\alpha}\left((\mu-1)(F{\cal P})_\alpha-\frac{\mu}{4}\partial_\alpha(JF)\right)\right].
\end{eqnarray}
The last term in (\ref{il3.11}) does not contributes to the equations of motion for $x,
{\cal P}$ and $J$, and can be omitted. Then the Hamiltonian for these variables acquires a simple and expected form
\begin{eqnarray}\label{phys.ham}
H=\frac{g_1}{2}\left({\cal{P}}^2-\frac{\mu e}{2c}(JF)+m^2c^2\right)\,.
\end{eqnarray}

The interaction yields two essential structural modifications of the theory.
Free theory implies the Frenkel condition, $J^{\mu\nu}\dot x_\nu=0$, and $p^\mu\sim\dot x^\mu$. Interaction modifies not only dynamical
equations but also the Frenkel condition, the latter necessarily turns into the Pirani condition
\begin{eqnarray}\label{uf4.7.4}
J^{\mu\nu}{\cal P}_\nu=0\,,
\end{eqnarray}
where, due to (\ref{uf4.7}), ${\cal P}^\mu$ is not proportional to $\dot x^\mu$. Then Eqs. (\ref{uf4.8}) and
(\ref{uf4.8-1}) imply that the interaction leads to a modification of the Lorentz-force equation even for uniform
fields. Only for the non anomalous value of magnetic moment, $\mu=1$, and uniform electromagnetic field the equations
(\ref{il2.5}) and (\ref{il2.6}) would be the same as in free theory, $\lambda_6=g_7=0$. Then
$T^{\mu\nu}=\eta^{\mu\nu}$, $Y^{\mu}=0$, and four-velocity becomes proportional to ${\cal P}^\mu$. Contribution of
anomalous magnetic moment $\mu\ne 1$ to the difference between $u$ and ${\cal P}$ is proportional to
$\frac{J}{c^3}\sim\frac{\hbar}{c^3}$, while the term with a gradient of field is proportional to
$\frac{J^2}{c^3}\sim\frac{\hbar^2}{c^3}$.

The remaining ambiguity due to $g_1$ in the equations (\ref{uf4.8})-(\ref{uf4.9}) reflects the reparametrization
symmetry of the theory. Assuming that the functions $x^\mu(\tau)$, $p^\mu(\tau)$ and $J^{\mu\nu}(\tau)$ represent the
physical variables $x^i(t)$, $p^\mu(t)$ and $J^{\mu\nu}(t)$ in the parametric form, their equations read
\begin{eqnarray}\label{uf4-ptp1}
\frac{dx^i}{dt} &=&c\frac{u^i}{u^0},\qquad \frac{dx^0}{dt}=c\,,\\
\label{uf4-ptp2} \frac{d {\cal P}^\mu}{dt} &=& \frac{e}{u^0}F^{\mu\nu} u_\nu+\frac{\mu e}{4u^0}\partial^\mu(JF)
\,,\\
\label{uf4-ptp3} \frac{dJ^{\mu\nu}}{dt}&=&\frac{c}{g_1u^0}\dot J^{\mu\nu}\,.
\end{eqnarray}
As it should be, they have unambiguous dynamics. Equations (\ref{uf4.8})- (\ref{uf4.9}) are written in an arbitrary
parametrization of the world-line. In the next subsection we exclude ${\cal P}^\mu$ and $g_1$, and then analyze the
resulting equations in the proper-time parameterizations. This allow us to compare them with original Frenkel
equations.

\subsection{Lagrangian form of equations \label{subsec:interactiion-L-eqs}}
Hamiltonian equations from the previous section can be rewritten in the Lagrangian form for the set $x, J$. Let us
analyze the relation between velocity and momentum given by the Hamiltonian equation (\ref{uf4.8}). This can be written
in the form
\begin{eqnarray}\label{pp1}
\dot x^\mu=g_1(T^\mu{}_\nu{\cal P}^\nu+Y^\mu)\,,
\end{eqnarray}
%
%
From this equation we express ${\cal P}$ through $\dot x$
\begin{eqnarray}\label{pp8}
{\cal P}^\mu=\frac{1}{g_1}\tilde T^\mu{}_\nu\dot x^\nu-\tilde T^\mu{}_\nu Y^\nu.
\end{eqnarray}
We can find $g_1$ calculating square of the following expression
\[
{\cal P}^\mu+\tilde T^\mu{}_\nu Y^\nu=\frac{1}{g_1}\tilde T^\mu{}_\nu\dot x^\nu\,,
\]
which yields
\[
{\cal P}^2+(\tilde T Y)^\mu(\tilde T Y)_\mu=\frac{1}{g_1^2}(\tilde T\dot x)^\mu(\tilde T\dot x)_\mu\,.
\]
We used that ${\cal P}_\mu\tilde T^\mu{}_\nu={\cal P}_\mu$ and ${\cal P}_\mu Y^\mu=0$. Using the last equation and
$T_1$\,-constraint we find $g_1$
\begin{eqnarray}\label{pp9}
g_1=\sqrt{\frac{(\tilde T\dot x)^2}{(\tilde T Y)^2-m^2c^2+\frac{\mu e(JF)}{2c}}}\equiv\frac{\sqrt{-g\dot x\dot
x}}{m_rc}\,,
\end{eqnarray}
where we have introduced the symmetric matrix
\begin{eqnarray}\label{pp10}
g_{\mu\nu}=(\tilde T^T\tilde T)_{\mu\nu}\,,
\end{eqnarray}
and the radiation mass
\begin{eqnarray}\label{pp10.1}
m_r^2=m^2-\frac{\mu e}{2c^3}(JF)-\frac{gYY}{c^2}\,.
\end{eqnarray}
In the natural parametrization $\sqrt{-g\dot x\dot x}=c$, we have $g_1=m_r^{-1}$, that is the auxiliary variable, which
appeared in front of mass-shell constraint $T_1=0$, is the inverse radiation mass. Due to the identity $(\tilde
TY)^\mu=\frac{b}{a}Y^\mu$ we also can write $gYY=\frac{b^2}{a^2}Y^2$. Using (\ref{pp8}) and (\ref{pp9}) in
(\ref{uf4.8-1}) and (\ref{uf4.9}) we write closed system of equations for $x^\mu$ and $J^{\mu\nu}$ in the form
\begin{eqnarray}\label{pp11}
\frac{d}{d\tau}\left[m_rc\frac{(\tilde T \dot x)^\mu}{\sqrt{-g\dot x\dot x}}-(\tilde T Y)^\mu\right]=\frac{e}{c}(F\dot
x)^\mu+ \frac{\mu e\sqrt{-g\dot x\dot x}}{4m_rc^2}\partial^\mu(JF)\,,
\end{eqnarray}
\begin{eqnarray}\label{pp11.1}
\dot J^{\mu\nu}=\frac{e\mu}{m_rc^2}\sqrt{-g\dot x\dot x}F^{[\mu}{}_\alpha
J^{\alpha\nu]}-\frac{2b(\mu-1)m_rc}{\sqrt{-g\dot x\dot x}}\dot x^{[\mu}(JF\dot x)^{\nu]}+\frac{2b}{a}\dot
x^{[\mu}Y^{\nu]}\,,
\end{eqnarray}
\begin{eqnarray}\label{pp11.2}
J^{\mu\nu}\tilde T_{\nu\alpha}(m_rc\dot x^\alpha-\sqrt{-g\dot x\dot x}Y^\alpha)=0\,.
\end{eqnarray}
Let us compare them with Frenkel equations. Frenkel found equations of motion consistent with the condition
$J^{\mu\nu}u_\nu=0$ up to order $O^3(J, F, \partial F)$. Besides, he considered the case $\mu=1$. Taking these
approximations in our equations in the proper-time parametrization $\sqrt{-(\dot x)^2}=c$, we arrive at those of
Frenkel (our J is $\frac{2mc}{e}$ of Frenkel $J$)
\begin{eqnarray}\label{pp11f}
\frac{d}{d\tau}\left[(m-\frac{e}{4mc^3}(JF))\dot
x^\mu+\frac{e}{8m^2c^3}J^{\mu\alpha}\partial_\alpha(JF)\right]=\frac{e}{c}(F\dot
x)^\mu+\frac{e}{4mc}\partial^\mu(JF)\,,
\end{eqnarray}
\begin{eqnarray}\label{pp11.1f}
\dot J^{\mu\nu} =\frac{e}{mc}\left[F^{[\mu}{}_\alpha J^{\alpha\nu]}-\frac{1}{4mc^2}\dot
x^{[\mu}J^{\nu]\alpha}\partial_\alpha(JF)\right]\,, \qquad J^{\mu\nu}\dot x_\nu=0\,.
\end{eqnarray}

In general case, our equations (\ref{pp11})-(\ref{pp11.2}) involve two types of corrections as compared with those of
Frenkel. First, the energy of magnetic moment in non uniform field leads to the contribution $-\frac{gYY}{c^2}$ into
the Frenkel radiation mass, see (\ref{pp10}). Second, when $\mu\ne 0$, a contribution arises because the Frenkel
condition which has been satisfied for the free particle, turns into Pirani condition in the interacting theory. Its
Lagrangian form is written in (\ref{pp11.2}). In the result, the components $J^{0i}$ vanish in the frame ${\cal
P}^\mu=({\cal P}^0, \vec 0)$ instead of the rest frame. Hence our model predicts small dipole electric moment of the
particle.

The structure of our equations simplified significantly for the stationary homogeneous field $\partial_\alpha
F^{\mu\nu}=0$. In this case (\ref{pp11}) and (\ref{pp11.1}) read
\begin{eqnarray}\label{pp11-F0}
\left[\frac{ m_r(\tilde T \dot x)^\mu}{\sqrt{-(g\dot x\dot x)}}\right]\dot{}=\frac{e}{c^2}(F\dot x)^\mu\,,
\end{eqnarray}
\begin{equation}
\label{pp11-J-F0} \dot J^{\mu\nu}=\frac{e\mu}{m_rc^2}\sqrt{-g\dot x\dot x}F^{[\mu}{}_\alpha
J^{\alpha\nu]}-2b\frac{(\mu-1)m_rc}{\sqrt{-g\dot x\dot x}}\dot x^{[\mu}(JF\dot x)^{\nu]}\,,
\end{equation}
\begin{eqnarray}\label{pp11.22}
(J\tilde T\dot x)^\mu=0\,.
\end{eqnarray}
Eq. (\ref{pp11-J-F0}) implies $\dot J^{\mu\nu}F_{\mu\nu}=0$. Hence $JF$ and $m_r$ are conserved quantities. Then
$T_1=0$ implies that ${\cal P}^2$ is a conserved quantity as well. The equation $\dot m_r=0$ can also be obtained
contracting (\ref{pp11-F0}) with $(\tilde T\dot x)_\mu$ and using the identity $v_\mu\left[\frac{
v^\mu}{\sqrt{-v^2}}\right]\dot{}\equiv 0$.

Contracting (\ref{pp11-F0}) with $T$ we can further simplify this equation
\begin{eqnarray}\label{pp15}
\frac{d}{d\tau}\left[\frac{m_r\dot x^\mu}{\sqrt{-g\dot x \dot x}}\right]=\frac{e}{c^2}(F'\dot x)^\mu, \qquad
F'=TF-\frac{m_rc^2}{e\sqrt{-g\dot x \dot x}}T\dot{\tilde T}.
\end{eqnarray}
%
Let us choose a parametrization which implies
\begin{eqnarray}\label{pp19}
g_{\mu\nu}\dot x^\mu\dot x^\nu=-c^2\,.
\end{eqnarray}
Since  $g\dot x\dot x=\dot x^2+O(J^2)$, in the linear approximation on $J$ this is just the proper-time
parametrization. Then the equations (\ref{pp11-J-F0}) and (\ref{pp15}) read
\begin{eqnarray}\label{pp20}
\frac{d(\tilde T \dot x)^\mu}{d\tau}=\frac{e}{m_rc}(F\dot x)^\mu\,,\quad {\rm or},\quad \ddot
x^\mu=\frac{e}{m_rc}(F'\dot x)^\mu, \qquad F'=TF-\frac{m_rc}{e}T\dot{\tilde T},
\end{eqnarray}
\begin{equation}
\label{pp19-J-F0} \dot J^{\mu\nu}=\frac{e\mu}{m_rc}F^{[\mu}{}_\alpha J^{\alpha\nu]}-2b(\mu-1)m_r\dot x^{[\mu}(JF\dot
x)^{\nu]}\,,
\end{equation}
So, when $\mu\ne 0$, the exact equations differ from the approximate equations (\ref{pp11f}) and (\ref{pp11.1f}) even
for uniform fields.

\subsection{BMT vector in Frenkel theory\label{sec:BMT}}
Since $J^{\mu\nu}{\cal P}_\nu=0$, the spin-tensor is equivalent to the four-vector (\ref{1.6}) where we replace
$p^\mu\rightarrow{\cal P}^\mu$. Then $s^\mu{\cal P}_\mu=0$. Due to Eqs. (\ref{uf4.7}) and (\ref{uf4.71}) together with
(\ref{1.4}), $s^\mu$ also obeys the condition
\begin{eqnarray}\label{uf4.121}
s^\mu u_\mu=s^\mu\dot x_\mu=0.
\end{eqnarray}
The physical dynamics can be described using $s^\mu$ instead of $J^{\mu\nu}$. Eq. (\ref{uf4.121}) suggests that $s^\mu$
could be candidate for BMT-vector in our model. Using the identities
\begin{eqnarray}\label{uf4.13}
J^{\mu\nu}=\frac{2}{\sqrt{-{\cal P}^2}}\epsilon^{\mu\nu\alpha\beta}s_\alpha {\cal P}_\beta, \qquad
\epsilon^{\mu\nu\alpha\beta}J_{\alpha\beta}=\frac{4}{\sqrt{-{\cal P}^2}}{\cal P}^{[\mu}s^{\nu]},
\end{eqnarray}
to represent $J^{\mu\nu}$ through $s^\mu$ in Eqs. (\ref{uf4.8})-(\ref{uf4.9}), we obtain the closed system of equations
for spin-plane invariant quantities
\begin{eqnarray}\label{uf4.14}
\dot x^\mu &=& g_1\left[{\cal P}^\mu-\frac{2(\mu-1)a}{\sqrt{-{\cal P}^2}}\epsilon^{\mu\nu\alpha\beta}(F{\cal P})_\nu
s_\alpha {\cal P}_\beta -\right.\cr &  &\left.\qquad\qquad\qquad\qquad\quad \frac{\mu a}{{\cal
P}^2}\epsilon^{\mu\alpha\gamma\delta} \epsilon^{\rho\beta\gamma'\delta'}s_\gamma {\cal P}_\delta s_{\gamma'} {\cal
P}_{\delta'} \partial_\alpha F_{\rho\beta}
\right], \\
\label{uf4.14-P} \dot{\cal P}^\mu &=& \frac{e}{c}(F\dot x)^\mu +g_1\frac{\mu e}{2c\sqrt{-{\cal
P}^2}}\epsilon_{\alpha\beta\gamma\rho}s^\gamma {\cal P}^\rho\partial^\mu F^{\alpha\beta}\,,
\\
\label{uf4.15} \dot s^\mu &=&g_1\frac{e\mu}{c}\left[(Fs)^\mu+\frac{1}{{\cal P}^2}(sF{\cal P}){\cal
P}^\mu\right]-\frac{1}{{\cal P}^2}(\dot{\cal P}s){\cal P}^\mu. ~
\end{eqnarray}
These equations valid for arbitrary electro-magnetic fields. Let us consider the case of uniform field discussed by
Bargmann Michel and Telegdi. Then we can compare these equations with BMT equations. First, we should exclude ${\cal P}$ and
$g_1$ from equations (\ref{uf4.14-P}) and (\ref{uf4.15}) using (\ref{uf4.14}). In contrast to (\ref{pp1}), where
$\dot{x}^\mu$ is a linear function of ${\cal P}^\mu$ and $J^{\mu\nu}$, in (\ref{uf4.14}) $\dot{x}^\mu$ is a non-linear
function of ${\cal P}^\mu$ and $s^\mu$. Inverse function which express ${\cal P}^\mu$ as a function of $\dot{x}^\mu$,
$s^\mu$ exists, though we can't find its explicit form even in the
case of uniform fields. Formally using (\ref{pp8}) and (\ref{pp9}) in the case of uniform fields, $\partial_\alpha
F^{\mu\nu}=0$, we get
\begin{eqnarray}\label{pp16}
\dot s^\mu=\sqrt{-g\dot x\dot x}\frac{e\mu}{m_rc^2}(Fs)^\mu- \cr \frac{e}{m_rc^2\sqrt{-g\dot x\dot
x}}\left[(\mu-1)(sF\dot x)+\mu b(sFJF\dot x)\right](\tilde T\dot x)^\mu\,, \qquad s\dot x=0\,.
\end{eqnarray}
Eq. (\ref{pp16}) contains $J$ but for weak fields the corresponding contribution can be neglected. In the uniform field
and in the parametrization (\ref{pp19}) we have Eq. (\ref{pp20}) for $x$ and
\begin{eqnarray}\label{pp21}
\dot s^\mu=\frac{e\mu}{m_rc}(Fs)^\mu-\frac{e}{m_rc^3}\left[(\mu-1)(sF\dot x)+\mu b(sFJF\dot x)\right](\tilde T\dot
x)^\mu\,.
\end{eqnarray}
This can be compared with BMT-equations
\begin{eqnarray}\label{pp22}
\ddot x^\mu=\frac{e}{mc}(F\dot x)^\mu,
\end{eqnarray}
\begin{eqnarray}\label{pp23}
\dot s^\mu=\frac{e\mu}{mc}(Fs)^\mu-\frac{e}{mc^3}(\mu-1)(sF\dot x)\dot x^\mu.
\end{eqnarray}
We can also introduce BMT-tensor dual to $s^\mu$
\[
J_{BMT}^{\mu\nu}=\frac{2}{c}\epsilon^{\mu\nu\alpha\beta}s_\alpha\dot x_\beta\,.
\]
Due to (\ref{pp23}) this obeys the equation
\begin{eqnarray}\label{pp22-J}
\dot J_{BMT}^{\mu\nu}= \frac{e}{mc} \left[\mu F^{[\mu}{}_\alpha J_{BMT}^{\alpha\nu]}- \frac{(\mu-1)}{c^2}(J_{BMT}F\dot
x)^{[\mu}\dot x^{\nu]}\right]\,.
\end{eqnarray}
This can be compared with (\ref{pp19-J-F0}).

Obtaining their equation (\ref{pp23}) in uniform field, Bargmann, Michel and Telegdi supposed that the motion of
particle (\ref{pp22}) is independent from the motion of spin. Besides they looked for the equation for $s^\mu$ linear
on $s$ and $F$. Obtaining Eqs. (\ref{pp20}) and (\ref{pp21}) we have not made any supposition of such a kind. Our
approach is based on the variational formulation which satisfies all the necessary symmetries. The exact equations
(\ref{pp20}) and (\ref{pp21}) involve two types of essential corrections as compared with BMT equations. First, an
energy of magnetic moment in electromagnetic fields leads to the radiation mass $m_r$. Second, anomalous magnetic
moment affects trajectory of a particle.

Neglecting non linear on $F$ and $s$ terms in our equations (\ref{pp20}), (\ref{pp21}), we obtain those of Bargmann,
Michel and Telegdi. The same holds if we take the proper-time parametrization, $\dot x^\mu\dot x_\mu=-c^2$, instead of
(\ref{pp19}).

\section{Exact solution in uniform magnetic field \label{sec:interaction-example}}

BMT equations give important information about spin kinematics of relativistic particles. Integrability of BMT
equations in the case of rather general electromagnetic backgrounds were studied in \cite{lobanov1999BMT-solutions}.
Solutions to BMT equations in a constant magnetic field can be associated with those of Dirac \cite{myagkii2001bmt}.

In this section we would like to study the behavior of our particle in a constant magnetic field. We take BMT vector of
our model as the basic quantity for description of spin, and compare dynamics of our and BMT models. So we take for the
analysis the closed system of equations (\ref{uf4.14}). BMT equations were derived as approximate equations describing
polarization effects in accelerators. They do not include possible influence of spin to the Lorentz force, and are
linear in spin, field and anomalous magnetic moment. Therefore, comparing our results with BMT we will take into
account this approximations.

Consider a particle with initial momentum ${\cal P}^\mu(0)$ and BMT spin $s^\mu(0)$ moving in the uniform magnetic
field directed along $z$-axis, ${\bf B}=B{\bf e}_z$, of a laboratory Cartesian coordinate system defined by an
orthonormal basis $({\bf e}_x,{\bf e}_y,{\bf e}_z)$. We already established that ${\cal P}^2$ and $(FJ)$ are integrals
of motion for uniform fields. In the case of uniform magnetic field we have
\[
(FJ)=4\gamma\left[({\bf B}{\bf s})-({\bf B}\boldsymbol{\beta})(\boldsymbol{\beta}{\bf s})\right] \,.
\]
Initial values ${\cal P}^\mu(0)$, $s^\mu(0)$ should satisfy to the constraints of our model ($T_1=0$, $s^2=3\hbar^2/4$, ${\cal P}s=0$).

Here and through the rest of this section we use following notations
\[
\gamma=\frac{{\cal P}^0}{\sqrt{-{\cal P}^2}}\,,\qquad \boldsymbol{\beta}=\frac{\vec{{\cal P}}}{{\cal P}^0}\,,
\]
in accordance with our construction of Lorentz invariant $SO(3)$ spin fiber bundle \cite{DPM1}. Here, $\gamma$ plays a
role of relativistic factor which in the limit of free electron reads $\gamma=(1-{\bf v}^2/c^2)^{-1/2}$. Denote by
$\beta$ module of vector $\boldsymbol{\beta}$. The quantity $a$ given in Eq. (\ref{uf4.71}) is also a constant, which
practically (for the magnetic fields smaller than Schwinger field) can be taken as $a\approx -\frac{e}{2m^2c^3}$.

The Hamiltonian equations of motions (\ref{uf4.14})-(\ref{uf4.15}) written in the parametrization of physical time read
\begin{eqnarray}\label{ex4-ptp1}
\frac{d{\bf x}}{dt} &=&c\frac{{\bf u}}{u^0},\qquad \frac{dx^0}{dt}=c\,,\\
\label{ex4-ptp2} \frac{d {\boldsymbol{\cal P}}}{dt} &=& \frac{e}{u^0}[{\bf u},{\bf B}]\,,\qquad \frac{d  {\cal
P}^0}{dt} = 0
\,,\\
\label{ex4-ptp3} \frac{d{\bf s}}{dt}&=&\frac{e\mu}{u^0}\left([{\bf s},{\bf B}]+\frac{1}{{\cal P}^2}({\bf
s},{\boldsymbol{\cal P}},{\bf B}){\boldsymbol{\cal P}}\right)-\frac{e}{{\cal
P}^2u^0}({\bf s},{\bf u},{\bf B}){\boldsymbol{\cal P}} \,,\\
\label{ex4-ptp4}
\frac{ds^{0}}{dt}&=&\frac{e{\cal P}^0}{u^0{\cal P}^2}\left[\mu({\bf s},{\boldsymbol{\cal P}},{\bf B})-({\bf s},{\bf u},{\bf B})\right]\,,\\
u^0&=&{\cal P}^0\left[1+2a(\mu-1)\gamma\left(\boldsymbol{\beta}^2({\bf B}{\bf s})-({\bf
B}\boldsymbol{\beta})(\boldsymbol{\beta}{\bf s})\right)\right]\,,
\\
{\bf u} &=&{\boldsymbol{\cal P}} \left[1+2a(\mu-1)\gamma\left(({\bf B}{\bf s})-({\bf
B}\boldsymbol{\beta})(\boldsymbol{\beta}{\bf s})\right)\right] -\frac{2a(\mu-1)}{\gamma}{\bf B}({\boldsymbol{\cal
P}}{\bf s})\,,
\end{eqnarray}
where $[{\bf s},{\bf B}]$ and $({\bf s},{\boldsymbol{\cal P}},{\bf B})$ mean the vector and mixed product of
3-dimensional vectors.

The velocity is not collinear to the momentum ${\cal P}^\mu$, its projection to the 3-hyperplane orthogonal to ${\cal
P}^\mu$ is proportional to anomalous magnetic moment. As a result, in general there is no common rest frame for
velocity and momentum.

From (\ref{ex4-ptp2})  follows that ${\cal P}^0={\rm const}$, hence ${\bf{\cal P}}^2={\rm const}$, $\gamma={\rm const}$
and $\boldsymbol{\beta}^2={\rm const}$. Another integral of motion is the projection of momentum to the magnetic field,
$({\boldsymbol{\cal P}}{\bf B})={\rm const}$. To simplify our calculations we assume without loosing generality that
initial vector of momentum is orthogonal to magnetic field, $({\boldsymbol{\cal P}}{\bf B})=0$. Indeed, other values of
$({\boldsymbol{\cal P}}{\bf B})$ can be obtained by boosts along ${\bf B}$ which do not modify electromagnetic tensor
(${\bf B}'={\bf B}$, ${\bf E}'={\bf E}=0$).

For the motion with momentum orthogonal to magnetic field we obtain the following system of equations
\begin{eqnarray}\label{eq-of-mot-vecP-phys-time}
\frac{d\boldsymbol{{\cal P}}}{dt} &=& \Omega_p[\boldsymbol{{\cal P}},{\bf e}_z]\,,\\
\frac{d{\bf x}}{dt}&=& c\left(\frac{\Omega_p}{eB}\boldsymbol{\cal P}-\frac{2a(\mu-1)\Omega_s}{\gamma \mu e}{\bf
e}_z(\boldsymbol{{\cal P}}{\bf s})\right)\,,
\\
\frac{dS^0}{dt} &=& \Omega'_{s}({\bf s},\boldsymbol{\beta},{\bf e}_z)
\,,\\
\label{eq-of-mot-vecS-phys-time} \frac{d{\bf s}}{dt} &=&
 \Omega_s [{\bf s},{\bf e}_z]+\Omega'_{s}  \boldsymbol{\beta}({\bf s},\boldsymbol{\beta},{\bf e}_z)\,,
\end{eqnarray}
where we use the following constants (frequencies)
\begin{eqnarray}\label{ex-frequencies1}
\Omega_p &=& \frac{e B (1+2a(\mu-1)\gamma({\bf B}{\bf s}))}{{\cal P}^0\left(1+2a(\mu-1)\gamma \beta^2({\bf B}{\bf s})\right)}\,,\\
\label{ex-frequencies2}
\Omega_s &=& \frac{\mu eB}{{\cal P}^0\left(1+2a(\mu-1)\gamma \beta^2({\bf B}{\bf s})\right)}\,,\\
\label{ex-frequencies3} \Omega'_s &=& \frac{e B {\cal P}^0(\mu-1-2a(\mu-1)\gamma({\bf B}{\bf s}))}{{\cal
P}^2\left(1+2a(\mu-1)\gamma \beta^2({\bf B}{\bf s})\right)}=\gamma^2(\Omega_s-\Omega_p) \,.
\end{eqnarray}
Multiplying (\ref{eq-of-mot-vecS-phys-time}) by ${\bf B}$ we find that $({\bf B}{\bf s})={\rm const}$ as it should be.
From (\ref{eq-of-mot-vecP-phys-time}) we find that the vector $\boldsymbol{\cal P}$ rotates with a constant circular
frequency $\Omega_p$ in the plane, orthogonal to magnetic field. In the orthonormal basis $({\bf e}_x,{\bf e}_y,{\bf
e}_z)$ solution for $\boldsymbol{\cal P}$ yields
\begin{equation}\label{soleq-of-mot-vecP-phys-time}
\boldsymbol{{\cal P}}=|\boldsymbol{{\cal P}}^{(0)}|\left({\bf e}_x\cos\left(\Omega_p t +\phi_p\right)+{\bf
e}_y\sin\left(\Omega_p t +\phi_p\right)\right)\,.
\end{equation}
For simplicity we choose ${\bf e}_x$ to provide $\phi_p=0$.

To solve (\ref{eq-of-mot-vecS-phys-time}) we first note that $s_z={\rm const}$. From constraints $s{\cal P}=0$,
$s^2=\frac{3}{4}\hbar^2$ one can see that the spin ${\bf s}$ for a particle with momentum $\boldsymbol{{\cal P}}={\cal
P}^0\boldsymbol{\beta}$ belong to the following ellipsoid
\begin{equation}\label{def-spin-ellipsoid}
s_i(\delta^{ij}-\beta^i\beta^j)s_j=\frac{3}{4}\hbar^2\,,
\end{equation}
which is obtained from a sphere with radius $\hbar\sqrt{3}/2$ by stretching in $\gamma$ times in the direction of
$\boldsymbol{\cal P}$. The main principal axis of this ellipsoid is always directed along $\boldsymbol{\cal P}$.
Therefore we write $s_x{\bf e}_x+s_y{\bf e}_y=s_1\boldsymbol{\tau}_1+s_2\boldsymbol{\tau}_2$ where the new basis is
\[
\boldsymbol{\tau}_1={\bf e}_x\cos(\Omega_p t)+{\bf e}_y\sin(\Omega_p t)\,, \quad \mbox{then} \quad \frac{d
\boldsymbol{\tau}_1}{dt}=\Omega_p \boldsymbol{\tau}_2\,,
\]
\[
\boldsymbol{\tau}_2=-{\bf e}_x\sin(\Omega_p t)+{\bf e}_y\cos(\Omega_p t ))\,, \quad \mbox{then} \quad
\frac{d\boldsymbol{\tau}_2}{dt}=-\Omega_p \boldsymbol{\tau}_1\,,
\]
\[
\boldsymbol{\tau}_3={\bf e}_z\,,
\]
which is determined by two principal axes of the ellipsoid. Note, that we just choose convenient variables to solve the
differential equation without transition to another reference frame. As a result we obtain simple equations
\begin{eqnarray}\label{eq-of-mot-vecS-rot-coord1}
\frac{ds_1}{dt} =- \Omega'_s s_2\,,\qquad \frac{ds_2}{dt} &=& \frac{\Omega'_s}{\gamma^2}s_1\,,
\end{eqnarray}
which describe evolution of spin on the ellipsoid (\ref{def-spin-ellipsoid}). Auxiliary radius vector
$s_1\boldsymbol{\tau}_1+s_2\boldsymbol{\tau}_2$ rotates  with circular frequency $\Omega'_s\gamma^{-1}$. Solutions to
Eqs. (\ref{eq-of-mot-vecS-rot-coord1}) read
\begin{eqnarray}\label{soleq-of-mot-vecS-rot-coord1}
s_1(t) = s^{(0)} \cos\left(\frac{\Omega'_s}{\gamma}t+\phi\right) \,,\qquad s_2(t) =
\frac{s^{(0)}}{\gamma}\sin\left(\frac{\Omega'_s}{\gamma}t+\phi\right) \,.
\end{eqnarray}
The radius vector $\vec{s}$ moves on the ellipse which is obtained as intersection of ellipsoid
(\ref{def-spin-ellipsoid})  and plane $s_z={\rm const}$. $s^{(0)}$ is nothing but the semi-major axis of this ellipse,
therefore it is restricted by the following interval $0\leq s^{(0)}\leq \gamma\sqrt{3}\hbar/2$.

In terms of initial variables spacial components of
4-vector $S^\mu$ evolves as follows
\[
{\bf s}(t)={\bf e}_x s^{(0)} \left[ \cos\left(\frac{\Omega'_s}{\gamma}t+\phi\right)\cos\left(\Omega_pt\right)-
\frac{1}{\gamma}\sin\left(\frac{\Omega'_s}{\gamma}t+\phi\right)\sin\left(\Omega_pt\right) \right]+
\]
\[
{\bf e}_y s^{(0)} \left[ \cos\left(\frac{\Omega'_s}{\gamma}t+\phi\right)\sin\left(\Omega_pt\right)+
\frac{1}{\gamma}\sin\left(\frac{\Omega'_s}{\gamma}t+\phi\right)\cos\left(\Omega_pt\right) \right]+{\bf e}_zs_z^{(0)}\,,
\]
where constants $s^{(0)}$, $s_z^{(0)}$, $\phi$ are restricted by (\ref{def-spin-ellipsoid}). Note that the angular
velocity of precession of vector ${\bf s}$ around ${\bf B}$ is time-dependent. Nevertheless, the helicity
$(\boldsymbol{\cal P}{\bf s})$ changes with the constant rate, $\Omega'_s/\gamma$. Indeed, $(\boldsymbol{\cal P}{\bf
s})=s^{(0)}|\boldsymbol{{\cal P}}^{(0)}|\cos\left(\frac{\Omega'_s}{\gamma}t+\phi\right)$.

Now we can substitute solutions ${\bf s}(t)$ and $\boldsymbol{{\cal P}}(t)$ into equation for $\vec{x}$
\[
\frac{d{\bf x}}{dt}=c|\boldsymbol{{\cal P}}^{(0)}|\left( \frac{\Omega_p }{eB}\left({\bf e}_x\cos(\Omega_pt)+{\bf
e}_y\sin(\Omega_p t)\right) -\frac{2a(\mu-1)\Omega_s}{\gamma \mu e }s^{(0)}{\bf e}_z
\cos\left(\frac{\Omega'_s}{\gamma}t+\phi\right)\right)\,.
\]
Integrating the last equation we find the trajectory
\begin{eqnarray}\label{sol-homB-trajectory}
{\bf x}(t) = {\bf x}_c+\qquad\qquad\qquad\qquad\qquad\qquad \qquad\qquad\qquad\cr
 c|\boldsymbol{{\cal P}}^{(0)}|\left(\frac{1}{eB}({\bf e}_x\sin(\Omega_p t)-{\bf e}_y\cos(\Omega_p t )) -
\frac{2a(\mu-1)\Omega_s}{\Omega'_s\mu e}s^{(0)}{\bf e}_z \sin\left(\frac{\Omega'_s}{\gamma}t+\phi\right)\right).
\end{eqnarray}
Trajectory represents sum of two motions: circular motion in the plane orthogonal to ${\bf B}$ and oscillations along
${\bf B}$. These oscillations accompany variations of the helicity. Vector ${\bf x}_c$ defines the center of circle.
The amplitude of oscillations along ${\bf B}$
\begin{eqnarray}\label{sol.1}
\Delta z=-\frac{2ca(\mu-1)\Omega_s}{\Omega'_s\mu e}|\boldsymbol{{\cal P}}^{(0)}|s^{(0)}\lesssim \beta \lambda_C\,,
\end{eqnarray}
less than the Compton wave-length. The trajectory of the
particle is shown in Figure 1.

\begin{figure}[t] \centering
        \includegraphics[width=\columnwidth]{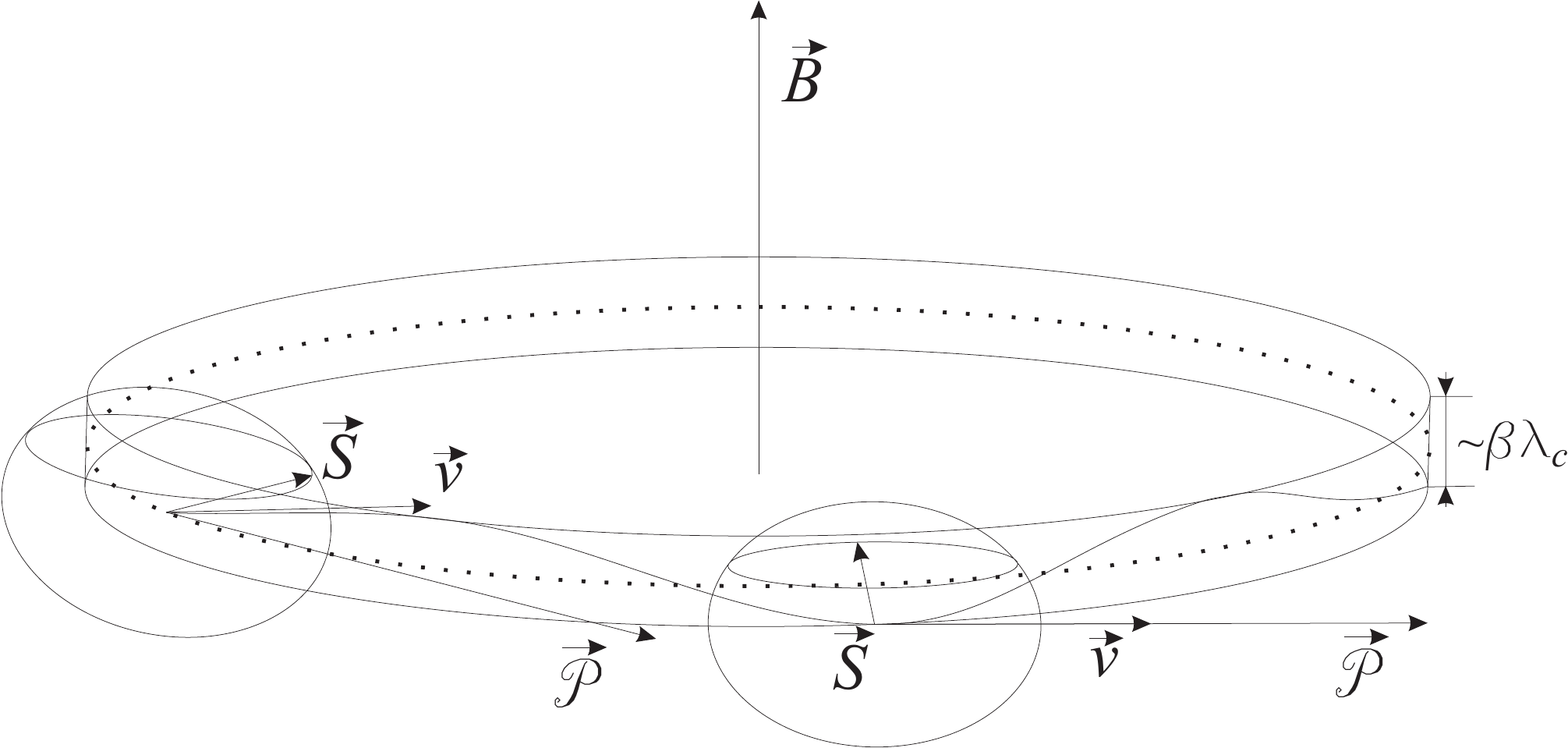}
    \caption{Momentum, velocity, spin and trajectory of a charged spinning particle in the uniform magnetic field}\label{fig:trajectory}
\end{figure}

Exact solution for this particular case demonstrates some important features of our model. There are two essentially
different situations with $\mu=1$ and $\mu\neq 1$. In the case of usual magnetic moment $\mu=1$, helicity is an
integral of motion, additional oscillations vanish and  the particle moves along circular trajectory in the plane
orthogonal to the magnetic field (dotted line in Figure \ref{fig:trajectory}).

In the case of anomalous magnetic moment, $\mu\ne 1$, helicity oscillates and affects the trajectory of particle. For a
small anomalous magnetic moment helicity and $z$-coordinate oscillate with a very slow rate. These oscillations of
trajectory along ${\bf B}$ with the amplitude of Compton wavelength can be called magnetic {\it Zitterbewegung}. When
$\mu\ne 1$, the velocity $\dot x^{\mu}$ and canonical momentum ${\cal P}_\mu$ are non-collinear. This seems to be usual
property of spinning particles \cite{corben:1968}. As we have started from Lagrangian variational problem, we have
explicit relation between velocity and canonical momentum. This allows us to exclude the canonical momentum from our
equations, see Eq. (\ref{pp20}). The result is an additional spin-orbit interaction, $F'\dot x$, instead of $F\dot x$.
Hence the magnetic {\it Zitterbewegung} appears due to the modification of Lorentz force for the spinning particle.

Magnetic {\it Zitterbewegung} leads to the corrections of the angular velocity of orbital motion $\Omega_p$ given by
\begin{eqnarray}\label{sol.2}
\Omega_p\approx\frac{e B}{\gamma mc} \left(1+\frac{e\gamma(\mu-1)}{ m^2c^3}({\bf s}{\bf B})+o(\hbar,\mu-1,B)\right)\,,
\end{eqnarray}
where $eB/(\gamma mc)$ is the angular velocity of spinless or BMT particle.
Frequency of helicity variations also corrected by high-order terms
\begin{eqnarray}\label{sol.3}
\frac{\Omega'_s}{\gamma}= (\mu-1)\frac{eB}{mc}\left(1-\frac{e\gamma}{m^2c^3}(1+(\mu-1)\beta^2)({\bf s}{\bf
B})+o(\hbar,\mu-1,B)\right) \,,
\end{eqnarray}
from the value $(\mu-1)\frac{eB}{mc}=(\frac{g}{2}-1)\frac{eB}{mc}$ computed by Bargmann, Michel and Telegdi \cite{BMT}.
The corrections are small, and for the experiments discussed by them, our equations give practically the same results.
Therefore our model is compatible with these experiments. Probably, other physical situations may be realized, where
the corrections could become notable. For instance, this may be the case of quasiparticles with large magnetic moment
\cite{karlovets2012}.

\section{Conclusions}
In this work we have presented solution to the problem which has been posed by Frenkel in 1926. He noticed that search
for variational formulation which takes into account the spin-tensor constraint $J^{\mu\nu}\dot x_\nu=0$ represents
rather non trivial problem. He found equations of motion consistent with this condition in the approximation $O^3(J, F,
\partial F)$, and when anomalous magnetic moment vanishes, $\mu=1$. We have found Lagrangian action
(\ref{lagrangian-bmt-em-4aux-vars}) for charged
spinning particle which implies all the desired constraints and equations of motion without approximations. They remain
consistent for any value of magnetic moment and for an arbitrary electromagnetic background. Besides, due to the
constraints (\ref{1.1})-(\ref{1.3}), our action guarantees the right number of both spacial and spin degrees of
freedom. In the above mentioned approximations, our equations coincide with those of Frenkel. In the recent work
\cite{DPW2}, we also demonstrated that the classical spinning particle has an expected behavior in arbitrary curved
background.

With the Lagrangian and Hamiltonian formulations at hands, we can unambiguously construct quantum mechanics of the
spinning particle and establish its relation with the Dirac equation. For the free theory, this has been done in the
work \cite{DPM2}. We showed that this gives
one-particle sector of the Dirac equation. Due to the second-class constraints (\ref{1.3}), the positions $x_i$ obey to
classical brackets with nonvanishing right hand side, see (\ref{xmxn-bmt-em-dirac67}). So, in the Dirac theory they realized
by non commutative operators which we identified with Pryce (d) center-of-mass \cite{pryce1948mass}. Since namely $x_i$
has an expected behavior (\ref{pp20}) as the position of spinning particle in classical interacting theory, our model
argue in favor of covariant Pryce (d) operator as the position operator of Dirac theory.

In resume, we have constructed variational formulation for relativistic spin one-half particle which is self consistent and
has reasonable behavior on both classical and quantum level.

As we have seen, interaction necessarily modifies some basic relations of the model. In the free theory the conjugated
momentum is proportional to velocity, $p^\mu\sim\dot x^\mu$ and the Frenkel condition holds. This is no more true in
interacting theory. The Frenkel condition turns into the Pirani condition, $J^{\mu\nu}{\cal P}_\nu=0$, where the
canonical momentum is not collinear to velocity. The advantage of Lagrangian formulation is that this gives exact
relation between them (see also Eqs. (\ref{pp1}) and (\ref{pp9})~)
\begin{eqnarray}\label{conc.1}
\dot x^\mu=g_1(T^\mu{}_\nu{\cal P}^\nu+Y^\mu)\,, \quad T^{\mu\nu}=\eta^{\mu\nu}+O(\mu-1, J), \quad Y^\mu=O(\partial
F, J)\,.
\end{eqnarray}
Only when $\mu=1$ and $\partial_\alpha F_{\mu\nu}=0$, the interacting and free theory have the same structure. To
resume, what happens in general case, let us consider our Hamiltonian equations with $Y^\mu=0$: $\dot
x^\mu=g_1T^\mu{}_\nu{\cal P}^\nu$, ~ $\dot{\cal P}^\mu=\frac{g_1e}{c}(F{\cal P})^\mu$, and compare them with the
standard expressions $\dot x^\mu=g_1{\cal P}^\mu$, ~ $\dot{\cal P}^\mu=\frac{g_1e}{c}(F{\cal P})^\mu$. Due to
$T^\mu{}_\nu$, excluding $\cal P$ from our equations, we obtain extra contributions to the standard expression for the
Lorentz force, $\ddot x^\mu=\frac{g_1e}{c}(F\dot x)^\mu+O(J)$. So the modification (\ref{conc.1}) means that complete
theory yields an extra spin-orbit interaction as compared with the approximate Frenkel and BMT equations.

We studied possible effects of this spin-orbit interaction in the case of uniform magnetic fields. The exact analytical
solution was obtained. Besides oscillations of the helicity first calculated by Bargmann, Michel and Telegdi, the
particle with anomalous magnetic moment experiences an effect of magnetic {\it Zitterbewegung} of the trajectory. Usual
circular motion in the plane orthogonal to ${\bf B}$ is perturbed by slow oscillations along ${\bf B}$ with the
amplitude of order of Compton wavelength. The Larmor frequency (\ref{sol.2}) and the frequency of  helicity
oscillations (\ref{sol.3}) are also shifted by small corrections. It would be  interesting to construct an experiment
which could detect these possible corrections, for instance due to resonance effects. Another possibility is an
artificial simulation of a point-like system with spin and a large anomalous magnetic moment. This could be inspired by
simulations of {\it Zitterbewegung} itself with a trapped ions \cite{Zahringer:2010}.

\section{Acknowledgments}
The work of AAD has been supported by the Brazilian foundation CNPq. AMPM thanks CAPES for the financial
support (programm PNPD/2011).

\appendix
\section{Dirac brackets}\label{sec:app2}
We construct Dirac brackets that take into account the second-class pairs $T_3$, $T_4$, $T_5$ and $T_6$. We will
calculate them iteratively in the case of arbitrary electromagnetic background. Then Dirac brackets of the free theory
can be obtained by substitution $F^{\mu\nu}=0$. We start from the pair of second class constraints is $T_6$ and $T_7$,
\[
\triangle_{67}=\{T_6,T_7\}={\cal P}^2+\frac{e}{4c}F_{\mu\nu}J^{\mu\nu}\,.\] At the constraint surface
$\triangle_{67}=-M^2c^2$.
The Poisson brackets of initial variables with constraints $T_6$ and $T_7$ are given in table
\ref{tabular:constraints34-variables-BMTem}.
\begin{table}
\caption{Constraints vs. variables} \label{tabular:constraints34-variables-BMTem}
\begin{center}
\begin{tabular}{c|c|c|c|c|c}
$\{,\}$                   & $x^\mu$      & $\mathcal{P}^\mu$      & $\pi^\mu$        & $\omega^\mu$      & $J^{\mu\nu}$ \\
\hline \hline $T_6=\mathcal{P}\omega$   &$-\omega^\mu$ & $-\frac{e}{c}F^{\mu\nu}\omega_\nu$ &$\mathcal{P}^\mu$ & 0 &
$2(\omega^\mu \mathcal{P}^\nu-\omega^\nu\mathcal{P}^\mu)$   \\
$T_7=\mathcal{P}\pi$      & $-\pi^\mu$   & $-\frac{e}{c}F^{\mu\nu}\pi_\nu$    & 0                &$-\mathcal{P}^\mu$ &
$2(\pi^\mu \mathcal{P}^\nu-\pi^\nu\mathcal{P}^\mu)$        \\
\end{tabular}
\end{center}
\end{table}
Using table \ref{tabular:constraints34-variables-BMTem} we calculate the Dirac brackets of basic variables with respect
to constraints $T_6$ and $T_7$
\[
\{Q_1,Q_2\}_{67}=\{Q_1,Q_2\}+\frac{1}{\triangle_{67}} \left(\{Q_1,T_6\}\{T_7,Q_2\}-\{Q_1,T_7\}\{T_6,Q_2\} \right).
\]
The brackets read
\begin{eqnarray}\label{xmxn-bmt-em-dirac67}
\{x^\mu,x^\nu\}_{67} &=& \frac{-J^{\mu\nu}}{2\triangle_{67}}\,,\\
\label{xmPn-bmt-em-dirac67}
\{x^\mu,\mathcal{P}^\nu\}_{67} &=& \eta^{\mu\nu}+\frac{e}{2c\triangle_{67}}J^{\mu\alpha}{F_\alpha}^\nu \equiv T^{\mu\nu}_{(0)}\,,\\
\label{xmwn-bmt-em-dirac67}
\{x^\mu,\omega^\nu\}_{67} &=& \frac{-\omega^{\mu}\mathcal{P}^\nu}{\triangle_{67}}\,,\\
\label{xmpin-bmt-em-dirac34}
\{x^\mu,\pi^\nu\}_{67} &=& \frac{-\pi^{\mu}\mathcal{P}^\nu}{\triangle_{67}}\,,\\
\label{xmJab-bmt-em-dirac34} \{x^\mu,J^{\alpha\beta}\}_{67} &=& \frac{1}{\triangle_{67}}\left(
J^{\mu\alpha}\mathcal{P}^\beta- J^{\mu\beta}\mathcal{P}^\alpha\right)\,,
\end{eqnarray}
\begin{eqnarray}
\label{wmpin-bmt-em-dirac67}
\{\omega^\mu,\omega^\nu\}_{67} &=& 0\,,\\
\label{pimpin-bmt-em-dirac67}
\{\pi^\mu,\pi^\nu\}_{67} &=& 0\,,\\
\label{wmpin-bmt-em-dirac67}
\{\omega^\mu,\pi^\nu\}_{67} &=& \eta^{\mu\nu}-\frac{\mathcal{P}^{\mu}\mathcal{P}^\nu}{\triangle_{67}}\equiv G^{\mu\nu}\,,\\
\label{wmJab-bmt-em-dirac67} \{\omega^\mu,J^{\alpha\beta}\}_{67} &=&
2(\omega^\alpha G^{\mu\beta}-\omega^\beta G^{\mu\alpha})\,,\\
\label{pimJab-bmt-em-dirac67} \{\pi^\mu,J^{\alpha\beta}\}_{67} &=&
2(\pi^\alpha G^{\mu\beta}-\pi^\beta G^{\mu\alpha})\,,\\
\label{JmnJab-bmt-em-dirac67} \{J^{\mu\nu},J^{\alpha\beta}\}_{67} &=& 2(G^{\mu\alpha} J^{\nu\beta}-G^{\mu\beta}
J^{\nu\alpha}-G^{\nu\alpha} J^{\mu\beta} +G^{\nu\beta} J^{\mu\alpha})\,,
\end{eqnarray}
\begin{eqnarray}\label{PmPn-bmt-em-dirac67}
\{\mathcal{P}^\mu,\mathcal{P}^\nu\}_{67} &=& \frac{e}{c}F^{\mu\nu}+\frac{e^2}{2\triangle_{67}c^2}(FJF)^{\mu\nu}
=\frac{e}{c}F^\mu{}_\alpha T^{\alpha\nu}_{(0)}\,,\\
\label{Pmwn-bmt-em-dirac67}
\{\mathcal{P}^\mu,\omega^\nu\}_{67} &=& -\frac{e}{\triangle_{67}c}F^{\mu\alpha}\omega_{\alpha}\mathcal{P}^\nu\,,\\
\label{Pmpin-bmt-em-dirac67}
\{\mathcal{P}^\mu,\pi^\nu\}_{67} &=& -\frac{e}{\triangle_{67}c}F^{\mu\alpha}\pi_{\alpha}\mathcal{P}^\nu\,,\\
\label{PmJab-bmt-em-dirac67} \{\mathcal{P}^\mu,J^{\alpha\beta}\}_{67} &=&
-\frac{e}{\triangle_{67}c}F^{\mu\nu}(\mathcal{P}^\alpha J_{\nu}{}^{\beta}-\mathcal{P}^\beta J_{\nu}{}^{\alpha})\,.
\end{eqnarray}

We have defined define tensor $G^{\mu\nu}$ as the Dirac bracket of spin variables $\omega^\mu$ and $\pi^\nu$. Besides,
$T^{\mu\nu}_{(0)}=T^{\mu\nu}(\mu=0)$, where $T^{\mu\nu}=\eta^{\mu\nu}-\frac{e(\mu-1)}{2c\triangle_{67}}(JF)^{\mu\nu}$.

On the next step we calculate Dirac brackets for the pair $\{T_4, T_5\}_{67}=2(T_4+a_4)$,
\[
\{Q_1,Q_2\}_{4567}=\{Q_1,Q_2\}_{67}+\frac{1}{2\omega^2}
\left(\{Q_1,T_4\}_{67}\{T_5,Q_2\}_{67}-\{Q_1,T_5\}_{67}\{T_4,Q_2\}_{67} \right).
\]
The Dirac brackets $\{,\}_{67}$ of initial variables with $T_4$ and $T_5$ are given in table
\ref{tabular:constraints56-variables-BMTem}.
\begin{table}
\caption{Constraints vs. variables} \label{tabular:constraints56-variables-BMTem}
\begin{center}
\begin{tabular}{c|c|c|c|c|c}
$\{,\}_{67}$       & $x^\mu$      & $\mathcal{P}^\mu$ & $\pi^\mu$     & $\omega^\mu$ & $J^{\mu\nu}$ \\
\hline \hline
$T_4=\omega^2-a_4$ & $0$          & $0$               & $2\omega^\mu$ &$0$           & $0$          \\
$T_5=\omega\pi$    &$0$           & $0$               &$\pi^\mu$      & $-\omega^\mu$& $0$          \\
\end{tabular}
\end{center}
\end{table}

From table \ref{tabular:constraints56-variables-BMTem} it is seen that variables $x^\mu, \mathcal{P}^\mu, J^{\mu\nu}$
have vanishing Dirac brackets $\{,\}_{67}$ with constraints $T_4$ and $T_5$.  Therefore, new Dirac brackets
$\{,\}_{4567}$  coincide with old Dirac brackets $\{,\}_{67}$ when at least one of arguments is a function
$Z(x^\mu,\mathcal{P}^\mu,J^{\mu\nu})$ of variables $x^\mu, \mathcal{P}^\mu$ and $J^{\mu\nu}$ only,
\begin{eqnarray}\label{db4567-db45}
\{Z,Q\}_{4567}=\{Z,Q\}_{67}\,, \qquad Z=Z(x^\mu,\mathcal{P}^\mu,J^{\mu\nu})\,.\\
\end{eqnarray}
We omit subscripts of brackets, so that $\{ ~ , ~ \}$ means $\{ ~ , ~ \}_{4567}$. The only modification in the Dirac
brackets comes from the basic variables in the spin sector
\begin{eqnarray}
\label{wmwn-bmt-em-dirac4567}
\{\omega^\mu,\omega^\nu\} &=& 0\,,\\
\label{wmpin-bmt-em-dirac4567} \{\omega^\mu,\pi^\nu\} &=&
\eta^{\mu\nu}-\frac{\mathcal{P}^{\mu}\mathcal{P}^\nu}{\triangle_{67}} -\frac{\omega^\mu \omega^\nu}{\omega^2}
\,,\\
\label{pimpin-bmt-em-dirac4567} \{\pi^\mu,\pi^\nu\} &=& \frac{-J^{\mu\nu}}{2\omega^2}
\,.\\
\end{eqnarray}
Thus the complete list of Dirac brackets $\{{}, {}\}$ consist of expressions
(\ref{xmxn-bmt-em-dirac67})-(\ref{xmJab-bmt-em-dirac34}), (\ref{wmwn-bmt-em-dirac4567})-(\ref{pimpin-bmt-em-dirac4567})
and (\ref{wmJab-bmt-em-dirac67})-(\ref{PmJab-bmt-em-dirac67}).

Now the second class constraints can be put equal to zero, therefore we can rewrite the Hamiltonian as follows
\begin{eqnarray}\label{phys.ham.1}
H=\frac{g_1}{2}({\cal P}^2-\frac{e\mu}{2c}(FJ)+m^2c^2)\,.
\end{eqnarray}
The constraint $T_3$ can also be excluded since it has zero Dirac brackets with $T_1$ and with all spin-plane invariant
variables of the theory. This Hamiltonian generates evolution
\[
\dot{x}^\mu=\{x^\mu,H\}\,,\qquad \dot{\cal P}^\mu=\{{\cal P}^\mu,H\}\,,\qquad \dot{J}^{\mu\nu}=\{{J}^{\mu\nu},H\}\,.
\]
To check consistency of our calculations, let us obtain equations of motion using the Dirac brackets $\{{}, {}\}$ (and
taking into account that at the constrained surface $\triangle_{67}=-M^2c^2=-m^2c^2-\frac{e(2\mu+1)}{4c^3}(FJ)~ $).

Equation for coordinate reads
\[
\dot{x}^\mu=\frac{g_1}{2}\{x^\mu,{\cal P}^2-\frac{e\mu}{2c}(FJ)\} =\]
\[
g_1\{x^\mu,{\cal P}^\nu\}{\cal P}_\nu -g_1\frac{e\mu}{4c}\{x^\mu,J^{\alpha\beta}\}F_{\alpha\beta}
-g_1\frac{e\mu}{4c}\{x^\mu,F^{\alpha\beta}\}J_{\alpha\beta} =\]
\[
g_1\left(\eta^{\mu\nu}-\frac{e}{2M^2c^3}J^{\mu\alpha}{F_\alpha}^\nu\right){\cal P}_\nu- \frac{g_1 e\mu}{4M^2c^3}
\left(\mathcal{P}^\alpha J^{\mu\beta}-\mathcal{P}^\beta J^{\mu\alpha}\right) F_{\alpha\beta} -\frac{g_1
e\mu}{8M^2c^3}J^{\mu\rho}\partial_\rho(FJ) =
\]
\[
g_1\left(\eta^{\mu\nu}-\frac{e}{2M^2c^3}J^{\mu\alpha}{F_\alpha}^\nu\right){\cal P}_\nu +g_1\frac{e\mu}{2M^2c^3}
J^{\mu\beta}F_{\beta\alpha}\mathcal{P}^\alpha -g_1\frac{e\mu}{8M^2c^3}J^{\mu\rho}\partial_\rho(FJ)=
\]
\[
g_1\left(\eta^{\mu\nu}+\frac{e(\mu-1)}{2M^2c^3}J^{\mu\alpha}{F_\alpha}^\nu\right){\cal P}_\nu-
g_1\frac{e\mu}{8M^2c^3}J^{\mu\rho}\partial_\rho(FJ)=g_1u^\mu\,,
\]
Equation for momentum reads
\[
\dot{\cal P}^\mu=\frac{g_1}{2}\{{\cal P}^\mu,{\cal P}^2-\frac{e\mu}{2c}(FJ)\}=
\]
\[
=g_1\{{\cal P}^\mu,{\cal P}^\nu\}{\cal P}_\nu -g_1\frac{e\mu}{4c}\{{\cal P}^\mu,J^{\alpha\beta}\}F_{\alpha\beta}
-g_1\frac{e\mu}{4c}\{{\cal P}^\mu,x^\rho\}\partial_\rho F^{\alpha\beta}J_{\alpha\beta} =\]
\[
=g_1\left(\frac{e}{c}F^{\mu\nu}+\frac{e^2}{2M^2c^4}F^{\mu\alpha}F^{\nu\beta}J_{\alpha\beta}\right){\cal P}_\nu
+g_1\frac{e^2\mu}{2M^2c^4}F^{\mu\nu}\mathcal{P}^\alpha {J^{\beta}}_{\nu}F_{\alpha\beta} +\]
\[g_1\frac{e\mu}{4c}
\left( \eta^{\mu\rho}-\frac{e}{2M^2c^3}J^{\rho\alpha}{F_\alpha}^\mu \right)
\partial_\rho (FJ)
=\]
\[
=g_1\frac{e}{c}F^{\mu\alpha}\left(\left(\eta_\alpha{}^\nu+\frac{e(\mu-1)}{2M^2c^3}J_{\alpha\beta}F^{\beta\nu}\right){\cal
P}_\nu -\frac{e\mu}{8M^2c^3} J_{\alpha\rho}
\partial^\rho (FJ)\right)
+g_1\frac{e\mu}{4c}
\partial^{\mu}(FJ)
=\]
\[
=g_1\frac{e}{c}F^{\mu\alpha}u_\alpha +g_1\frac{e\mu}{4c}
\partial^{\mu}(FJ)\,,
\]
Equation for spin-tensor reads
\[
\dot{J}^{\alpha\beta}=\frac{g_1}{2}\{{J}^{\alpha\beta},{\cal P}^2-\frac{e\mu}{2c}(FJ)\}=
\]
\[
=g_1\{{J}^{\alpha\beta},{\cal P}^\mu\}{\cal P}_\mu -g_1\frac{e\mu}{4c}\{{J}^{\alpha\beta},J^{\mu\nu}\}F_{\mu\nu}
-g_1\frac{e\mu}{4c}\{{J}^{\alpha\beta},x^\mu\}\partial_\mu (FJ) =\]
\[
=g_1\frac{e}{M^2c^3}F^{\mu\nu}(\mathcal{P}^\alpha {J^{\beta}}_{\nu}-\mathcal{P}^\beta {J^{\alpha}}_{\nu}){\cal P}_\mu
+g_1\frac{e\mu}{c}(G^{\mu\alpha} J^{\nu\beta}-G^{\mu\beta} J^{\nu\alpha})F_{\mu\nu} +\]
\[g_1\frac{e\mu}{4c}\frac{1}{M^2c^2}\left(\mathcal{P}^\alpha J^{\mu\beta}-\mathcal{P}^\beta J^{\mu\alpha}\right)\partial_\mu (FJ)
=\]
\[
=-g_1\frac{e}{M^2c^3}\mathcal{P}^{\left[\alpha\right.} {J^{\left.\beta\right]}}_{\nu} F^{\nu\mu}{\cal P}_\mu
+g_1\frac{e\mu}{c}(F^{\alpha}{}_{\nu} J^{\nu\beta}-F^{\beta}{}_{\nu} J^{\nu\alpha})
+g_1\frac{e\mu}{M^2c^3}\mathcal{P}^{\left[\alpha\right.} {J^{\left.\beta\right]}}_{\nu}F^{\nu\mu}{\cal P}_\mu +\]
\[g_1\frac{e\mu}{4c}\frac{1}{M^2c^2}\left(\mathcal{P}^\alpha J^{\mu\beta}-\mathcal{P}^\beta J^{\mu\alpha}\right)\partial_\mu (FJ)
=\]
\[
=g_1\frac{e}{c}\left(\mu(F^{\alpha}{}_{\nu} J^{\nu\beta}-F^{\beta}{}_{\nu} J^{\nu\alpha})
+\frac{(\mu-1)}{M^2c^2}\mathcal{P}^{\left[\alpha\right.} {J^{\left.\beta\right]}}_{\nu} F^{\nu\mu}{\cal P}_\mu -
\frac{1}{4M^2c^2}\mathcal{P}^{\left[\alpha\right.} {J^{\left.\beta\right]}}_{\mu}\partial^\mu (FJ)\right)\,.
\]

The Hamiltonian is proportional to the first class constraint $T_1$. Therefore equations of motion contain arbitrary
function $g_1(\tau)$ which is related with reparametrization invariance of the model. To obtain unambiguous equations
of evolution we can impose the gauge $x^0=c\tau$. The gauge is often called canonical gauge. Constraint $T_1$ together
with this condition form a pair of second class constraints. We have
\[
\{x^0-c\tau,T_1\}=2u^0\,,
\]
and Dirac brackets in the canonical gauge read
\[
\{Q_1,Q_2\}_{\tau}=\{Q_1,Q_2\}+\frac{1}{2u^0} \left(\{Q_1,G_1\}\{T_1,Q_2\}-\{Q_1,T_1\}\{G_1,Q_2\}\right).
\]

The Dirac brackets of constraint $T_1$ and canonical gauge $G_1$ with physical variables are given in table
\ref{tabular:canonical-gauge-variables-BMTem}. There compact notations
\[\dot{\cal P}^{\mu }\equiv\frac{g_1}{2}\{{\cal P}^{\mu },T_1\}\,,\]
\[
\dot{J}^{\mu\nu}\equiv\frac{g_1}{2}\{{J}^{\mu\nu},T_1\}\,,
\]
\[
u^\mu\equiv\left(\eta^{\mu\nu}+\frac{e(\mu-1)}{2M^2c^3}J^{\mu\alpha}{F_\alpha}^\nu\right){\cal P}_\nu-
\frac{e\mu}{8M^2c^3}J^{\mu\rho}\partial_\rho(FJ)\,.
\]
\[
C=-\frac{e}{c}(\mu-1)(\omega F{\cal P})+\frac{e\mu}{4c}(\omega\partial)(JF)\,,
\]
\[
D=-\frac{e}{c}(\mu-1)(\pi F{\cal P})+\frac{e\mu}{4c}(\pi\partial)(JF)\,,
\]
were used.

\begin{table}
\caption{Constraints vs. variables (canonical gauge)} \label{tabular:canonical-gauge-variables-BMTem}
\begin{center}
\begin{tabular}{c|c|c|c|c|c}
$\{,\}_{4567}$                         & $x^\mu$                      & $\mathcal{P}^\mu$  & $\omega^\mu$ & $\pi^\mu$        &  $J^{\mu\nu}$ \\
\hline \hline $T_1={\cal P}^2-\frac{e\mu}{2c}(FJ)$ & $-2u^\mu$                     & $\frac{-2}{g_1}\dot{\cal P}^{\mu
}$ &
$\frac{2C{\cal P}^\mu}{\triangle_{67}}-\frac{2e\mu}{c}(F\omega)^\mu$ & $\frac{2D{\cal P}^\mu}{\triangle_{67}}-\frac{2e\mu}{c}(F\pi)^\mu$&  $\frac{-2}{g_1}\dot{J}^{\mu\nu}$       \\
$G_1=x^0-c\tau$                       & $\frac{-1}{2\triangle_{67}}J^{0\mu}$  & $T^{0\mu}_{(0)}$   & $\frac{-\omega^0{\cal P}^\mu}{\triangle_{67}}$ & $\frac{-\pi^0{\cal P}^\mu}{\triangle_{67}}$                  &  $\frac{-1}{\triangle_{67}}J^{0\left[\nu\right.}{\cal P}^{\left.\mu\right]}$ \\
\end{tabular}
\end{center}
\end{table}
The Dirac brackets which take into account the canonical gauge are as follow. \par \noindent Spacial sector:
\begin{eqnarray}\label{xmxn-bmt-em-dirac-canonical-gauge}
\{x^\mu,x^\nu\}_{\tau} &=& \frac{-1}{2u^0\triangle_{67}}\left(u^0J^{\mu\nu}-u^\mu J^{0\nu}+u^\nu J^{0\mu}\right)\,,\\ \nonumber
\label{xmPn-bmt-em-dirac-canonical-gauge} \{x^\mu,\mathcal{P}^\nu\}_{\tau} &=&
\eta^{\mu\nu}-\frac{u^\mu}{u^0}\eta^{0\nu}+\frac{e}{2u^0c\triangle_{67}} \left(
u^0J^{\mu\alpha}-u^\mu J^{0\alpha}+ u^\alpha J^{0\mu}\right){F_\alpha}^\nu \\ \nonumber &-&
\frac{e\mu}{8u^0c\triangle_{67}}J^{0\mu}\partial^\nu(FJ)
 \,,\\ \nonumber
\label{PmPn-bmt-em-dirac-canonical-gauge} \{\mathcal{P}^\mu,\mathcal{P}^\nu\}_{\tau} &=&\frac{e}{u^0c} \left(
u^0F^\mu{}_\alpha T^{\alpha\nu}_{(0)}-\frac{c}{eg_1}\dot{\cal P}^{\mu}T^{0\nu}_{(0)} +\frac{c}{eg_1}\dot{\cal P}^{\mu}
T^{0\nu}_{(0)} \right) \,.
\end{eqnarray}
Frenkel sector:
\begin{eqnarray}
\label{JmnJab-bmt-em-dirac-canonical-gauge} \{J^{\mu\nu},J^{\alpha\beta}\}_{\tau} &=& \{J^{\mu\nu},J^{\alpha\beta}\}
-\frac{1}{g_1u^0\triangle_{67}} \left( \dot{J}^{\mu\nu} J^{0\left[\beta\right.}{\cal P}^{\left.\alpha\right]}-
\dot{J}^{\alpha\beta} J^{0\left[\nu\right.}{\cal P}^{\left.\mu\right]}\right)\,, \\ \nonumber
\label{xmJab-bmt-em-dirac-canonical-gauge} \{x^\mu,J^{\alpha\beta}\}_{\tau} &=& \frac{-1}{u^0\triangle_{67}}\left(
u^\mu J^{0\left[\alpha\right.}{\cal P}^{\left.\beta\right]} -u^0 J^{\mu\left[\alpha\right.}{\cal
P}^{\left.\beta\right]}\right)-\frac{1}{2u^0\triangle_{67}g_1}J^{0\mu}\dot{J}^{\alpha\beta}\,, \\ \nonumber
\label{PmJab-bmt-em-dirac-canonical-gauge} \{\mathcal{P}^\mu,J^{\alpha\beta}\}_{\tau} &=&
\frac{e}{u^0c\triangle_{67}}F^{\mu}{}_{\nu} \left( u^0\mathcal{P}^{\left[\alpha\right.}
J^{\left.\beta\right]\nu}-u^\nu\mathcal{P}^{\left[\beta\right.} {J^{\left.\alpha\right] 0}}
\right)-\frac{1}{u^0g_1}T^{0\mu}_{(0)}\dot{J}^{\alpha\beta} \\ \nonumber &-&
\frac{e\mu}{4u^0c\triangle_{67}}\partial^{\mu}(FJ)\mathcal{P}^{\left[\beta\right.} J^{\left.\alpha\right] 0}  \,.
\end{eqnarray}
Basic spin variables:
\begin{eqnarray}
\label{wmwn-bmt-em-dirac-canonical-gauge}
\{\omega^\mu,\omega^\nu\}_{\tau} &=& -\frac{e\mu \omega^0}{2u^0c\triangle_{67}}(F^{\mu\alpha}{\cal P}^\nu-
F^{\nu\alpha}{\cal P}^\mu)\omega_\alpha\,,\\ \nonumber
\label{wmpin-bmt-em-dirac-canonical-gauge} \{\omega^\mu,\pi^\nu\}_{\tau} &=&
\eta^{\mu\nu}-\frac{\mathcal{P}^{\mu}\mathcal{P}^\nu}{\triangle_{67}}\left(1+\frac{(\pi^0C+\omega^0D)}{u^0c\triangle_{67}}\right)
-\frac{\omega^\mu \omega^\nu}{\omega^2} \\ \nonumber &-&
\frac{e\mu \omega^0}{u^0c\triangle_{67}}\left(\pi^0F^{\mu\alpha}\omega_\alpha{\cal
P}^\nu-\omega^0F^{\nu\alpha}\pi_\alpha{\cal P}^\mu\right)\,,
\\ \nonumber
\label{pimpin-bmt-em-dirac-canonical-gauge} \{\pi^\mu,\pi^\nu\}_{\tau} &=& \frac{-J^{\mu\nu}}{2\omega^2}-\frac{e\mu
\pi^0}{2u^0c\triangle_{67}}(F^{\mu\alpha}{\cal P}^\nu-F^{\nu\alpha}{\cal P}^\mu)\pi_\alpha \,.
\end{eqnarray}
Other mixed brackets:
\begin{eqnarray}
\label{xmwn-bmt-em-dirac-canonical-gauge} \{x^\mu,\omega^\nu\}_{\tau} &=&
\frac{-\omega^{\mu}\mathcal{P}^\nu}{\triangle_{67}} +\frac{1}{u^0}
\left(\frac{J^{0\mu}}{2\triangle_{67}}\left(\frac{C{\cal P}^\nu}{\triangle_{67}}-\frac{e\mu}{c}(F\omega)^\nu\right)+
\frac{u^\mu\omega^0{\cal P}^\nu}{\triangle_{67}}\right)
\,,\\ \nonumber
\label{xmpin-bmt-em-dirac-canonical-gauge} \{x^\mu,\pi^\nu\}_{\tau} &=&
\frac{-\pi^{\mu}\mathcal{P}^\nu}{\triangle_{67}} +\frac{1}{u^0}
\left(\frac{J^{0\mu}}{2\triangle_{67}}\left(\frac{D{\cal P}^\nu}{\triangle_{67}}-
\frac{e\mu}{c}(F\pi)^\nu\right)+ \frac{u^\mu\pi^0{\cal P}^\nu}{\triangle_{67}}\right)\,,\\ \nonumber
\label{wmJab-bmt-em-dirac-canonical-gauge} \{\omega^\mu,J^{\alpha\beta}\}_{\tau} &=&
2(\omega^\alpha G^{\mu\beta}-\omega^\beta G^{\mu\alpha}) \\ \nonumber &-&
\frac{1}{u^0\triangle_{67}} \left(\omega^0{\cal P}^\mu\frac{\dot{J}^{\alpha\beta}}{g_1}+\left(\frac{C{\cal
P}^\mu}{\triangle_{67}}-\frac{e\mu}{c}(F\omega)^\mu\right)J^{0\left[\beta\right.}{\cal P}^{\left.\alpha\right]}\right)
\,,\\ \nonumber
\label{pimJab-bmt-em-dirac-canonical-gauge} \{\pi^\mu,J^{\alpha\beta}\}_{\tau} &=& 2(\pi^\alpha G^{\mu\beta}-\pi^\beta
G^{\mu\alpha}) \\ \nonumber &-&
\frac{1}{u^0\triangle_{67}} \left(\pi^0{\cal P}^\mu\frac{\dot{J}^{\alpha\beta}}{g_1}+\left(\frac{D{\cal
P}^\mu}{\triangle_{67}}-\frac{e\mu}{c}(F\pi)^\mu\right)J^{0\left[\beta\right.}{\cal P}^{\left.\alpha\right]}\right) \,,
\\ \nonumber
\label{Pmwn-bmt-em-dirac-canonical-gauge}
\{\mathcal{P}^\mu,\omega^\nu\}_{\tau} &=& -\frac{e}{\triangle_{67}c}F^{\mu\alpha}\omega_{\alpha}\mathcal{P}^\nu
\\ \nonumber &-&
\frac{1}{u^0} \left(T^{0\mu}_{(0)}\left(\frac{C{\cal P}^\nu}{\triangle_{67}}-\frac{e\mu}{c}(F\omega)^\nu\right)-
\frac{\dot{\cal P}^{\mu}}{g_1}\frac{\omega^0{\cal P}^\nu}{\triangle_{67}}\right)\,,\\ \nonumber
\label{Pmpin-bmt-em-dirac-canonical-gauge}
\{\mathcal{P}^\mu,\pi^\nu\}_{\tau} &=& -\frac{e}{\triangle_{67}c}F^{\mu\alpha}\pi_{\alpha}\mathcal{P}^\nu
\\ \nonumber &-&
\frac{1}{u^0} \left(T^{0\mu}_{(0)}\left(\frac{D{\cal P}^\nu}{\triangle_{67}}-\frac{e\mu}{c}(F\pi)^\nu\right)-
\frac{\dot{\cal P}^{\mu}}{g_1}\frac{\pi^0{\cal P}^\nu}{\triangle_{67}}\right)\,.
\end{eqnarray}

In the free theory the algebra of Dirac brackets simplifies significantly. In this case ${\cal P}^\mu=p^\mu$,
$u^\mu=p^\mu$, $\dot{J}^{\mu\nu}=\dot{\cal P}^\mu=0$, $\triangle_{67}=p^2$, and in an arbitrary parametrization $\tau$,
we have the following brackets:
\par
\noindent Basic variables of spin:
\begin{eqnarray}\label{db1}
\{\omega^\mu,\omega^\nu\}=0, \quad \{\omega^\mu,\pi^\nu\}=g^{\mu\nu}-\frac{\omega^\mu\omega^\nu}{\omega^2}, \quad
\{\pi^\mu,\pi^\nu\}=-\frac{1}{2\omega^2}J^{\mu\nu};
\end{eqnarray}
\begin{eqnarray}\label{db2}
\{x^\mu,\omega^\nu\}=-\frac{\omega^{\mu}p^\nu}{p^2}, \qquad \{x^\mu,\pi^\nu\}=-\frac{\pi^{\mu}p^\nu}{p^2};
\end{eqnarray}
Spacial sector:
\begin{equation}\label{db3}
\{x^\mu,x^\nu\}=-\frac{1}{2p^2}J^{\mu\nu}\,,\qquad \{x^\mu,p^\nu\}=\eta^{\mu\nu}, \qquad \{p^\mu,p^\nu\}=0.
\end{equation}
Frenkel sector:
\begin{equation}\label{db5}
\{J^{\mu\nu},J^{\alpha\beta}\}= 2(g^{\mu\alpha} J^{\nu\beta}-g^{\mu\beta} J^{\nu\alpha}-g^{\nu\alpha} J^{\mu\beta}
+g^{\nu\beta} J^{\mu\alpha})\,,
\end{equation}
\begin{equation}\label{db6}
\{x^\mu,J^{\alpha\beta}\}=\frac{1}{p^2}(J^{\mu\alpha}p^\beta- J^{\mu\beta}p^\alpha )\,,
\end{equation}
BMT-sector: take $s^\mu=\frac{1}{4\sqrt{-p^2}}\epsilon^{\mu\nu\alpha\beta}p_\nu J_{\alpha\beta}$, then
\begin{eqnarray}\label{db7}
\{s^\mu,s^\nu\}=-\frac{1}{\sqrt{-p^2}}\epsilon^{\mu\nu\alpha\beta}p_\alpha s_\beta=\frac{1}{2}J^{\mu\nu},
\end{eqnarray}
\begin{eqnarray}\label{db8}
\{x^\mu,s^\nu\}=-\frac{s^\mu p^\nu}{p^2}=-\frac{1}{4\sqrt{-p^2}}\epsilon^{\mu\nu\alpha\beta}J_{\alpha\beta}-\frac{p^\mu
s^\nu}{p^2}.
\end{eqnarray}
Other Dirac brackets vanish. In the equations (\ref{db1}) and (\ref{db5}) it has been denoted
\begin{equation}\label{db9}
g^\mu{}_\nu\equiv\delta^\mu{}_\nu-\frac{p^{\mu}p_\nu}{p^2}\,.
\end{equation}
Together with $\tilde g^\mu{}_\nu\equiv\frac{p^\mu p_\nu}{p^2}$, this forms a pair of projectors
\begin{equation}\label{db10}
g+\tilde g=1, \quad g^2=g, \quad \tilde g^2=\tilde g, \quad g\tilde g=0\,.
\end{equation}

The free Dirac brackets which take into account the canonical gauge are as follows.
\par
\noindent Basic variables of spin:
\begin{eqnarray}\label{wmwn-bmt-free-dirac-canonical-gauge}
\{\omega^\mu,\omega^\nu\}_{\tau}=0, \quad
\{\omega^\mu,\pi^\nu\}_{\tau}=g^{\mu\nu}-\frac{\omega^\mu\omega^\nu}{\omega^2}, \quad
\{\pi^\mu,\pi^\nu\}_{\tau}=-\frac{1}{2\omega^2}J^{\mu\nu}\,,
\end{eqnarray}
\begin{eqnarray}
\label{xmwn-bmt-free-dirac-canonical-gauge} \{x^\mu,\omega^\nu\}_{\tau} &=& \frac{-\omega^{\mu}{p}^\nu}{p^2}
+\frac{\omega^0p^\mu{p}^\nu}{p^0p^2}
\,,\\
\label{xmpin-bmt-free-dirac-canonical-gauge} \{x^\mu,\pi^\nu\}_{\tau} &=& \frac{-\pi^{\mu}{p}^\nu}{p^2}
+\frac{\pi^0p^\mu{p}^\nu}{p^0p^2}\,,
\end{eqnarray}
Spacial sector:
\begin{eqnarray}\label{xmxn-bmt-free-dirac-canonical-gauge}
\{x^\mu,x^\nu\}_{\tau} &=& \frac{-1}{2p^0p^2}\left(p^0J^{\mu\nu}-p^\mu J^{0\nu}+p^\nu J^{0\mu}\right)\,,\\
\label{xmPn-bmt-free-dirac-canonical-gauge} \{x^\mu,p^\nu\}_{\tau} &=& \eta^{\mu\nu}-\frac{p^\mu}{p^0}\eta^{0\nu}\,,
\qquad \{p^\mu,p^\nu\}=0\,.
\end{eqnarray}
Frenkel sector:
\begin{equation}\label{db5}
\{J^{\mu\nu},J^{\alpha\beta}\}_\tau= 2(g^{\mu\alpha} J^{\nu\beta}-g^{\mu\beta} J^{\nu\alpha}-g^{\nu\alpha} J^{\mu\beta}
+g^{\nu\beta} J^{\mu\alpha})\,,
\end{equation}
\begin{equation}
\label{xmJab-bmt-free-dirac-canonical-gauge} \{x^\mu,J^{\alpha\beta}\}_{\tau} = \frac{-1}{p^0p^2}\left( p^\mu
J^{0\left[\alpha\right.}{p}^{\left.\beta\right]} -p^0 J^{\mu\left[\alpha\right.}{p}^{\left.\beta\right]}\right)\,,
\end{equation}
BMT-sector:
\begin{eqnarray}\label{SS-bmt-free-dirac-canonical-gauge}
\{s^\mu,s^\nu\}_\tau=-\frac{1}{\sqrt{-p^2}}\epsilon^{\mu\nu\alpha\beta}p_\alpha s_\beta=\frac{1}{2}J^{\mu\nu},
\end{eqnarray}
\begin{eqnarray}\label{SS-bmt-free-dirac-canonical-gauge}
\{x^\mu,s^\nu\}_\tau=-\frac{1}{4\sqrt{-p^2}}\epsilon^{\mu\nu\alpha\beta}J_{\alpha\beta}+
\frac{p^\mu}{4p^0\sqrt{-p^2}}\epsilon^{0\nu\alpha\beta}J_{\alpha\beta} =\frac{(s^0p^\mu p^\nu -s^\mu p^\nu
p^0)}{p^0p^2}.
\end{eqnarray}
Other Dirac brackets vanish.

Here we define Dirac brackets for all phase space variables. After transition to the Dirac brackets the second-class
constraints can be used as strong equalities, therefore it is enough to consider Dirac brackets at the constraint
surface only. Then, explicit form of the Dirac brackets depends on the choice of independent variables. For instance,
in the free theory considered in the gauge of physical time we can present $s_{BMT}^0$ and $p^0$ in terms of
independent variables ${\bf s}_{BMT}$, ${\bf p}$, ${\bf x}$
\begin{eqnarray}\label{def:S0-P0-canonical-gauge}
s^0=\frac{(\bf{s}\, \bf{p})}{\sqrt{{\bf p}^2+(mc)^2}}\,,\qquad p^0=\sqrt{{\bf p}^2+(mc)^2}\,.
\end{eqnarray}
The non vanishing Dirac brackets are
\begin{equation}\label{pha.15}
\{x^i,x^j\}_{\tau}=\frac{\epsilon^{ijk}s_k}{mcp^0}\,,\qquad \{x^i,p^j\}_{\tau}=\delta^{ij}\,, \qquad
\{p^i,p^j\}_{\tau}=0,
\end{equation}
\begin{eqnarray}\label{pha.16}
\{s^i,s^j\}_\tau=\frac{p^0}{mc}\epsilon^{ijk}\left(s_k- \frac{({\bf{s}\,\bf{p}})p_k}{p_0^2}\right)\,,
\end{eqnarray}
\begin{eqnarray}\label{pha.17}
\{x^i,s^j\}_D=\left(s^i-\frac{({\bf{s}\, \bf{p}})p^i}{p_0^2}\right)\frac{p^j}{(mc)^2}\,.
\end{eqnarray}

\end{document}